\documentclass[aps,pre,reprint,superscriptaddress,showpacs,showkeys,amsmath,amssymb,hidelinks,nofootinbib]{revtex4-1}
\usepackage[separate-uncertainty = true]{siunitx}
\usepackage{graphicx}
\usepackage{dcolumn}
\usepackage{bm}
\usepackage{hyperref}
\usepackage{etoolbox}
\usepackage{multirow, booktabs}
\usepackage[dvipsnames]{xcolor}
\makeatletter

\renewcommand*{\p@subsection}{}

\renewcommand*{\p@subsubsection}{}

\DeclareSIUnit\pixel{px}
\makeatother
\begin{document}
\preprint{v1.0}

\title{Deep neural networks for\\classifying complex features in diffraction images}

\author{Julian Zimmermann}
\email{julian.zimmermann@mbi-berlin.de}
\affiliation{Max-Born-Institut f\"ur Nichtlineare Optik und Kurzzeitspektroskopie, 12489 Berlin, Germany}%

\author{Bruno Langbehn}
\affiliation{Institut f\"ur Optik und Atomare Physik, Technische Universit\"at Berlin, 10623 Berlin, Germany}%

\author{Riccardo Cucini}
\affiliation{Elettra—Sincrotrone Trieste S.C.p.A., 34149 Trieste, Italy}%

\author{Michele Di Fraia}
\affiliation{Elettra—Sincrotrone Trieste S.C.p.A., 34149 Trieste, Italy}
\affiliation{ISM-CNR, Istituto di Struttura della Materia, LD2 Unit, 34149 Trieste, Italy}

\author{Paola Finetti}
\affiliation{Elettra—Sincrotrone Trieste S.C.p.A., 34149 Trieste, Italy}%

\author{Aaron C. LaForge}
\affiliation{Physikalisches Institut, Universit\"at Freiburg, 79104 Freiburg, Germany}

\author{Toshiyuki Nishiyama}
\affiliation{Division of Physics and Astronomy, Graduate School of Science, Kyoto University, Kyoto 606-8502, Japan}

\author{Yevheniy Ovcharenko}
\affiliation{Institut f\"ur Optik und Atomare Physik, Technische Universit\"at Berlin, 10623 Berlin, Germany}%
\affiliation{European XFEL GmbH, 22869 Schenefeld, Germany}

\author{Paolo Piseri}
\affiliation{CIMAINA and Dipartimento di Fisica, Università degli Studi di Milano, 20133 Milano, Italy}

\author{Oksana Plekan}
\affiliation{Elettra—Sincrotrone Trieste S.C.p.A., 34149 Trieste, Italy}%

\author{Kevin C. Prince}
\affiliation{Elettra—Sincrotrone Trieste S.C.p.A., 34149 Trieste, Italy}%
\affiliation{Department of Chemistry and Biotechnology, Swinburne University of Technology, Victoria 3122, Australia}

\author{Frank Stienkemeier}
\affiliation{Physikalisches Institut, Universit\"at Freiburg, 79104 Freiburg, Germany}

\author{Kiyoshi Ueda}
\affiliation{Institute of Multidisciplinary Research for Advanced Materials, Tohoku University, Sendai 980-8577, Japan}

\author{Carlo Callegari}
\affiliation{Elettra—Sincrotrone Trieste S.C.p.A., 34149 Trieste, Italy}
\affiliation{ISM-CNR, Istituto di Struttura della Materia, LD2 Unit, 34149 Trieste, Italy}

\author{Thomas M\"oller}
\affiliation{Institut f\"ur Optik und Atomare Physik, Technische Universit\"at Berlin, 10623 Berlin, Germany}%

\author{Daniela Rupp}
\affiliation{Max-Born-Institut f\"ur Nichtlineare Optik und Kurzzeitspektroskopie, 12489 Berlin, Germany}%

\date{\today}

\begin{abstract}
 Intense short-wavelength pulses from free-electron lasers and high-harmonic-generation sources enable diffractive imaging of individual nano-sized objects with a single x-ray laser shot. The enormous data sets with up to several million diffraction patterns represent a severe problem for data analysis, due to the high dimensionality of imaging data. Feature recognition and selection is a crucial step to reduce the dimensionality. Usually, custom-made algorithms are developed at a considerable effort to approximate the particular features connected to an individual specimen, but facing different experimental conditions, these approaches do not generalize well. On the other hand, deep neural networks are the principal instrument for today's revolution in automated image recognition, a development that has not been adapted to its full potential for data analysis in science. We recently published in [\citeauthor{Langbehn2018} Phys. Rev. Lett. 121, 255301 (2018)] the first application of a deep neural network as a feature extractor for wide-angle diffraction images of helium nanodroplets. Here we present the setup, our modifications and the training process of the deep neural network for diffraction image classification and its systematic benchmarking. We find that deep neural networks significantly outperform previous attempts for sorting and classifying complex diffraction patterns and are a significant improvement for the much-needed assistance during post-processing of large amounts of experimental coherent diffraction imaging data.
\end{abstract}

\maketitle

\section{\label{sec:introduction}Introduction}

Coherent diffraction imaging (CDI) experiments of single particles in free flight have been proven to be a significant asset in the pursuit of understanding the structural composition of nano-scaled matter \cite{Seibert2011, Loh2012, Bostedt2010, Gomez2014,Chapman2010, Rupp2017}. While traditional microscopy methods are able to image fixated, substrate-grown or deposited individual particles \cite{Li2008,Farges1986, Clemmer1997, Kostko2007, Sakdinawat2010}, only CDI can combine high-resolution images with single particles in free flight in one experiment \cite{Bostedt2009, Gorkhover2012, Bostedt2016}. CDI became possible due to the recent advent of short wavelength free-electron lasers (FELs) producing coherent high-intensity x-ray pulses with femtosecond duration with a single x-ray laser shot \cite{Emma2010}. However, CDI also comes with its own set of new challenges.

One of the growing problems of CDI experiments is the sheer amount of recorded data that has to be analyzed. 
The LINAC Coherent Light Source (LCLS), for instance, has a repetition rate of \SI{120}{\hertz} with typical hit-rates ranging from \SIrange{1}{30}{\percent} \cite{Emma2010,Bostedt2016a,Calvey2016}, greatly depending on the performed experiment.
The newly opened European XFEL will have an even higher maximum repetition rate of \SI{27000}{\hertz} \cite{Schneidmiller2011}, which may add up to several million diffraction patterns in a single \num{12}-hour shift.
The idea of using neural networks for classification of large number of scattering patterns was born out of the significant difficulties of analyzing large data sets of clusters \cite{Rupp2014}, in particularly metal clusters \cite{Barke2015}. Moreover, the ability to analyze such data sets is sought after by the community in general \cite{Lundholm2018}.
For example, for the successful determination of 3D-structures from a CDI data set using the expansion-maximization-compression algorithm \cite{Flamant2016,Ekeberg2015a,Lundholm2018}, it is necessary to sample the 3D Fourier space up to the Nyquist rate for the desired resolution and this for all sub-species contained in the target under study. The achievable resolution, as well as the chance for successful convergence of the algorithm, correlates directly with the number of diffraction patterns with a high signal-to-noise ratio \cite{Flamant2016}. Thus, huge data sets are taken and as a consequence of the sheer amount of data, it is getting increasingly complicated to distill the high-quality data subsets that are suitable for subsequent analysis steps.

The enormous success of neural networks in the regime of image processing and classification provides a unique way of facing the imminent data-analysis bottleneck and reduces the impending problem to a mere domain adaptation from datasets used throughout the industry to ones that are used in CDI research. This work aims to be a stepping-stone towards this adaptation by providing an introduction to the theory of deep neural networks and analyzing how to best transfer and optimize these algorithms to the domain of scattering images. As a new baseline, we train a widely used deep neural network architecture, a residual convolutional deep neural network \cite{He2016a}, in a supervised manner with a training set of manually labeled data. We then adapt the neural network to the domain of diffraction images and improve on the baseline performance by addressing the following issues:
\begin{enumerate}
 \item Modification of the architecture to account for the specificities of diffraction images and thus optimize the prediction capabilities.
 \item Determination of the appropriate size of the training dataset in order to keep the manual work of a researcher to a moderate level.
 \item Mitigation of experimental artifacts, in particular noisy diffraction images.
\end{enumerate}


\begin{figure}
 \includegraphics[width=1\columnwidth]{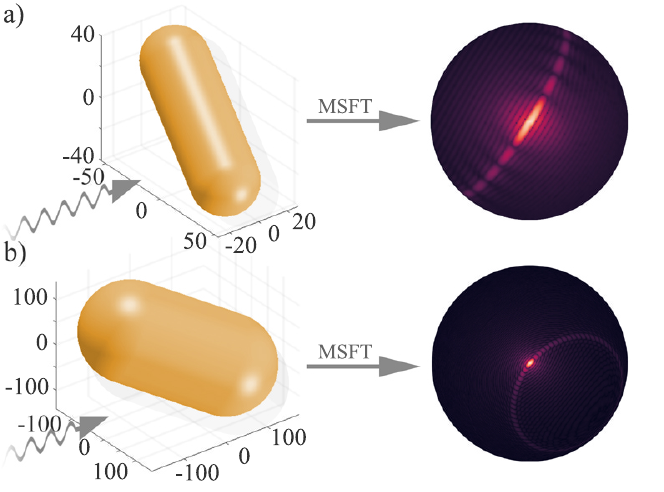}
 \caption{\label{fig:same_particle_different_pattern} a) and b) are showing a \emph{capsule} shaped particles whose orientation and size differs. The scattering images are calculated using a multi-slice Fourier transform (MSFT) algorithm that simulates a wide-angle x-ray scattering experiment which includes 3D information about the particle \cite{Barke2015, Rupp2017}. The two incoming beams (indicated by the arrow on the left-hand side) produce very different scattering images, yet the dominant feature, an elongated bent streak, is distinctly visible in both calculations. A handcrafted algorithm is typically not able to identify the similarity between the two scattering patterns and would classify these two images in two distinct classes, although they belong to the same \emph{capsule} shape class. A deep neural network can learn these complicated similarities on its own when we provide a few manually selected diffraction patterns that contain this feature.}
\end{figure}

Experience has shown that a researcher is able to relate diffraction patterns produced by similarly shaped particles of different sizes and orientations in context with each other. However, a programmatic description for a classification and sorting of these mostly similar patterns is almost impossible to achieve.

Figure \ref{fig:same_particle_different_pattern} illustrates the case of two diffraction patterns captured from almost identical particles but under different orientations. Both patterns clearly show an elongated and bent streak, but the bending is differently pronounced and directed. If we wanted to handcraft an algorithm that detects this feature, we would need to describe it via some appropriate metric that must take into account the various grades of inflection, direction, brightness, and completeness of this feature within every image.
Furthermore, we would need to redo it for every characteristic feature in a diffraction image of which we want to find similar ones.

In addition to that, poor signal-to-noise ratios, stray-light, a beam stop or central hole of multichannel plates or pnCCDs \cite{Meidinger2006} and overall poor image quality can even further increase the difficulty to make an automatized classification of all images coherent \cite{Kurta2016, Bobkov2015, Atla2011}.

Therefore, we need a robust classification routine that is insusceptible to the described artifacts, just as a researcher is, to tackle the upcoming data volume. Deep neural networks provide a way out of this situation, and we show in this paper that they outperform the current state-of-the-art classification and sorting routines.

Current state-of-the-art automatic classification routines for diffraction experiments employ so-called kernel methods \cite{Bobkov2015,Yoon2011}. \citet{Bobkov2015} trained a support-vector-machine on a public small-angle x-ray scattering dataset with an \emph{Accuracy} of \SI{87}{\percent}, but only on selected images (we will use this approach as a reference in section 4). \citet{Yoon2011} were able to achieve an \emph{Accuracy} of up to \SI{90}{\percent} using unsupervised spectral clustering on a non-public small-angle x-ray scattering dataset.

Deep neural networks, on the other hand, have already been applied to a broad range of physics-related problems ranging from predicting topological ground states \cite{Deng2017}, distinguish different topological phases of topological band insulators \cite{Zhang2018}, enhancing the signal-to-noise at hadron colliders \cite{Field1996}, differentiate between so-called known-physics background and new-physics signals at the Large Hadron Collider \cite{Bhimji2017} and to help solve the Schr\"odinger equation \cite{Mills2017,Manzhos2009}. Their ability to classify images has also been utilized in cryo-electron microscopy \cite{Zhu2017a}, medical imaging \cite{Gao2017a} and even for hit-finding in serial x-ray crystallography \cite{Ke2018a}. However, to our knowledge, this paper is the first application of deep neural networks for classifying complex features within diffraction patterns.
We show that deep neural networks outperform the current state-of-the-art classification and sorting routines, while being insusceptible to typical artifact features of diffraction measurements. Furthermore a deeper analysis of the trained network shows that it can understand complex concepts of what constitutes a characteristic feature in a diffraction pattern.

The paper is organized as follows: In section \ref{sec:dataset}, the data set is presented and a few experimental details are discussed.
Section \ref{sec:basic_theory} provides the fundamental theory to understand the basics of neural networks; it has two subsections.
Subsection \ref{subsec:deep} covers the theory, and algorithmic underpinnings of deep neural networks and how to train these models and subsection \ref{subsec:evaluating_neural_network} presents three common metrics to evaluate the quality of the neural network's predictions.

Section \ref{sec:deep neural network_baseline} establishes our starting point, while the full benchmark report on the baseline neural network can be found in appendix \ref{subsec:architecture_choices}. We introduce the chosen network architecture and provide baseline results on the data presented in section \ref{sec:dataset} but also on a reference dataset for which classification results are already published \cite{Kassemeyer2012}.

In section \ref{sec:deep neural network_for_scattering}, we discuss solutions for the above stated issues of applying neural networks to diffraction data.
In subsection \ref{subsec:activation_function} we discuss the choice of the activation function for the neural network and present a novel logarithmic activation function that enhances the prediction performance with diffraction image data.
Subsection \ref{subsec:size_of_training_set} benchmarks the dependence of neural networks on training data size, asking essentially how much manually labeled data is needed for the neural network to give acceptable results and subsection \ref{subsec:ccf} presents an approach to harden the neural network against very noisy data using a custom two-point cross-correlation map.

In section \ref{sec:interpretation_heatmaps} we then provide more profound insights into the output of the neural network by showing and discussing calculated heatmaps that visualize the gradient flow within the neural network. These images directly correlate with what the neural network \emph{sees}; they are created using an advanced visualization algorithm called GradCam++ \cite{Chattopadhyay2017}.

Finally, we give a summary of the principal results and unique propositions of this paper and conclude with an outlook on further modifications as well as future directions.


\section{\label{sec:dataset}The data}
Helium nanodroplets \cite{Langbehn2018} were imaged using extreme ultraviolet (XUV) photon energies between \SIrange{19}{35}{\electronvolt} using the experimental setup of the LDM beamline \cite{Lyamayev2013, Svetina2015} at the Free Electron Laser FERMI \cite{Allaria2012}.
Scattering images were recorded with a multi-channel-plate (MCP) detector combined with a phosphor screen which was placed \SI{65}{\milli\metre} downstream from the interaction region; this defines the maximum scattering angle of \SI{30}{\degree}. Single shot diffraction images in the XUV regime are in some respect a special case, as they cover large scattering angles and can contain 3D structural information \cite{Barke2015}, manifesting as complex and pronounced characteristic features, such as the bent streaks in Figure \ref{fig:same_particle_different_pattern}.
Out of \num{2e5} laser shots, about \num{38000} images were obtained. The images were corrected for straylight background and the flat detector (see also \citet{Langbehn2018})

\begin{figure*}
 \includegraphics[width=1\textwidth]{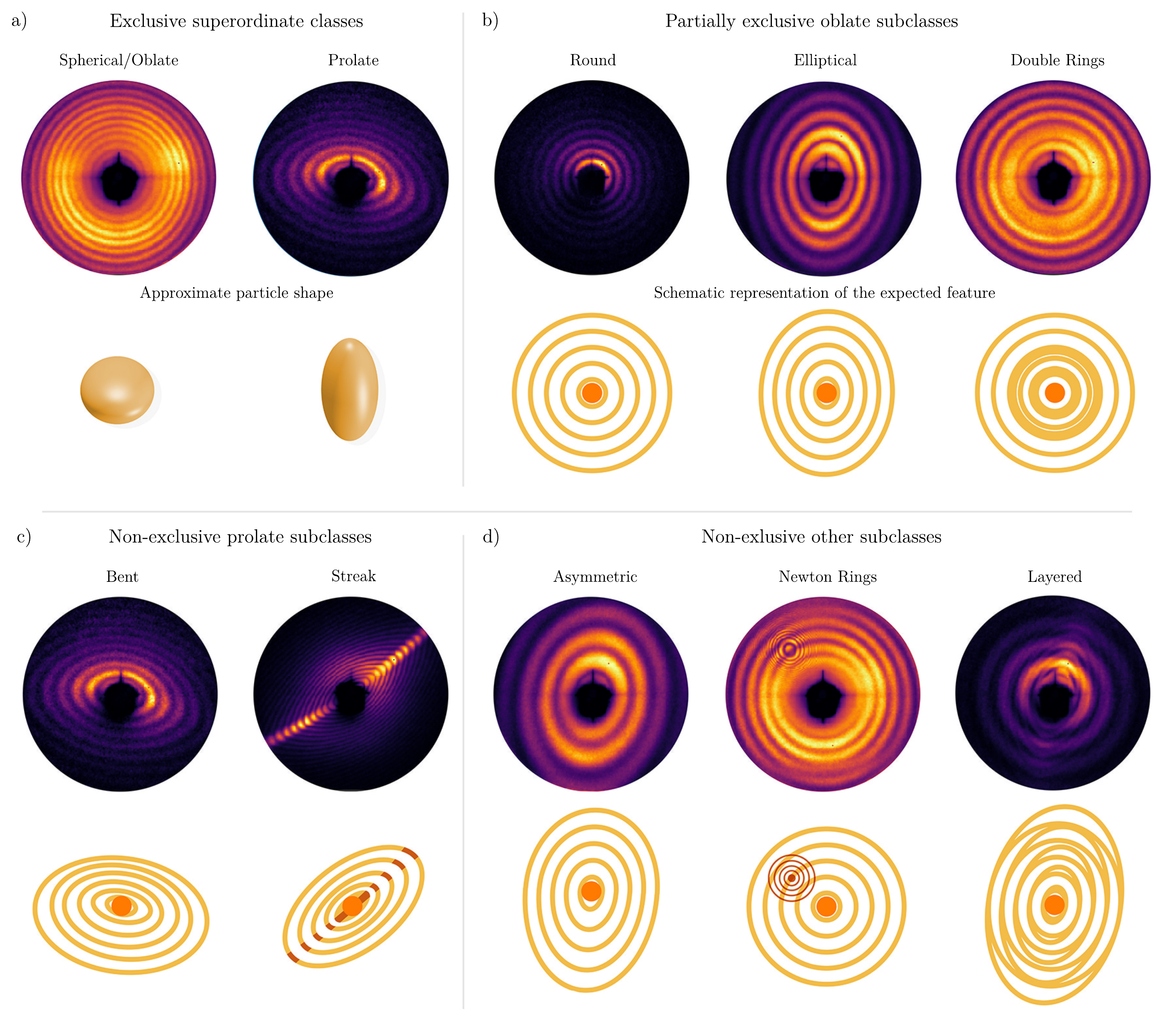}%
 \caption{\label{fig:helium_samples} Characteristic examples for all the classes assigned to the \num{7264} images by a researcher, except for the \emph{Empty} class. The top row of every class shows a representative diffraction pattern and the bottom row in b) - d) shows a stylized drawing of the characteristic feature of this class. The bottom row in a) shows an illustration of the name-giving particle shape for the \emph{Spherical/Oblate} and \emph{Prolate} class.
  The shapes are derived from the analysis of the data in \citet{Langbehn2018}, and they serve as a form of superordinate classes. They are mutually exclusive to each other, and all diffraction patterns are part of one of these two classes. Also, both superordinate classes have subclasses. For example, b) is showing the \emph{Spherical/Oblate} subclasses \emph{Round}, \emph{Elliptical} and \emph{Double Rings}. While a diffraction pattern can be part of the \emph{Round} and the \emph{Double Rings} class, it cannot be part of the \emph{Round} and the \emph{Elliptical} class. For the \emph{Prolate} superordinate class, we find analog subclass rules, although there is no exclusivity rule as it was with the \emph{Round} and \emph{Elliptical} class. Therefore, an image belonging to \emph{Bent} can also be in the \emph{Streaks} class.
  Furthermore, all \emph{Spherical/Oblate} and \emph{Prolate} patterns can not only be part of their respective subclass but can also be part of one or more of the classes in the \emph{non-exclusive other subclass} categories - shown in d). These classes describe general features within the image which are to some extent independent of the particle shape. We derived the superordinate classes from these general features.
  These complicated inter-class relationships demonstrate the capabilities of a researcher to interconnect mostly distinctive appearing features into a consistent description and ultimately leading to a valid physical interpretation. A hand-crafted algorithm could not account for these relationships normally, but now these interconnections can serve as an additional evaluation metric for the neural network. Since there is no diffraction pattern which belongs to the \emph{Spherical/Oblate} and \emph{Prolate} class simultaneously, we can check if the neural network mislabeled a diffraction pattern according to these rules.
  We can then interpret this as a reliable indicator for a failed generalization of the network. The physics behind these patterns are quite complicated as well, but for a rigorous interpretation and analysis of these patterns, please see \citet{Langbehn2018}.}
\end{figure*}

For the neural network training dataset, we selected \num{7264} diffraction images randomly out of all recorded patterns. The size of the subset was chosen to be the maximum a researcher could classify manually given one week time. From this subset we manually identified \num{11} distinct but non-exclusive classes (see Figure \ref{fig:helium_samples} for examples as well as a description and Table \ref{tab:helium_stats} for statistics about every class). We chose each of the diffraction patterns shown in Figure \ref{fig:helium_samples} for being a strong candidate for its class, but it is important to note that almost all diffraction patterns belong to multiple classes since this is a multi-class labeling scenario. These patterns are therefore not always clearly distinguishable from each other and can exhibit multiple characteristics from different classes. For example, the Newton rings in Figure \ref{fig:helium_samples}d) are superimposed on a concentric ring pattern that falls into the category \emph{Spherical/Oblate}, but Newton rings can also occur in other classes, e.g. streak patterns.
Furthermore, labeling all images is itself prone to systematic errors because the researcher has to learn-to-label \cite{Frenay2014}. This means that the labeling process itself is to some extent ill-posed, as the researcher does not know the characteristics of a feature a priori which results in a changing perception of features and classes along the labeling process and thus a systematically decreased consistency for every class.

We uploaded all available data alongside our assigned labels to the public CXI database (CXIDB, \cite{Maia2012}) under the public domain CC0 waiver \footnote{\url{http://cxidb.org/id-94.html}}.

\begin{table}[b]
 \caption{\label{tab:helium_stats}%
  Statistics of the helium nanodroplets dataset. Non-exclusive labels assigned by a researcher. One image can be in multiple classes. Total dataset size is \num{7264}. Note that \emph{Spherical/Oblate} as a class also contains \emph{Round} patterns, only \emph{Prolate} shapes are excluded from this class (see also caption of Figure \ref{fig:helium_samples}).
 }
 \begin{ruledtabular}
  \begin{tabular}{lcc}
   Class                     & Nr. of labels & \% of the whole dataset \\
   \colrule
   \textrm{Spherical/Oblate} & \num{6589}    & \num{90.7}              \\
   \textrm{Round}            & \num{5792}    & \num{79.7}              \\
   \textrm{Elliptical}       & \num{796}     & \num{11.0}              \\
   \textrm{Newton rings}     & \num{460}     & \num{6.3}               \\
   \textrm{Prolate}          & \num{453}     & \num{6.2}               \\
   \textrm{Bent}             & \num{390}     & \num{5.4}               \\
   \textrm{Asymmetric}       & \num{367}     & \num{5.1}               \\
   \textrm{Streak}           & \num{242}     & \num{3.3}               \\
   \textrm{Double Rings}     & \num{218}     & \num{3.0}               \\
   \textrm{Layered}          & \num{47}      & \num{0.7}               \\
   \textrm{Empty}            & \num{222}     & \num{3.1}               \\
  \end{tabular}
 \end{ruledtabular}
\end{table}

\section{\label{sec:basic_theory}Basic Theory}

\subsection{\label{subsec:deep}What is a deep neural network}
We concentrate in this paper solely on deep feed-forward neural networks. They are a classification model consisting of a directed acyclic graph that defines a set of hierarchically structured non-linear functions.

A fundamental example can be constructed by arranging \emph{n} non-linear functions ($z_{1},~z_{2},~\dots ~z_{n}$) in a chain-like manner:
$z_{\text{output}} = z_{n} \left( z_{n-1} \left( \dots \left( z_{2} \left( z_{1} \left(\bm{x} \right) \right) \right) \dots \right) \right)$, where
$\bm{x}$ is the input, which is in our case a diffraction image. The first function, $z_{1} \left(\bm{x} \right)$, is called the input layer.
We then pass the output of $z_{1}$ to $z_{2}$ and so on; this goes on until the last layer ($z_{n}$) which is called the output layer. The nomenclature is that all layers except the output layer ($z_{n}$) and the input layer ($z_{1}$) are called hidden layers.

For illustrative purposes, Figure \ref{fig:nn_schematic} shows a convolutional neural network. There, we schematically show the layer functions $z_{1},~\dots ~z_{n}$ where every layer consists of two stages; A linear layer-specific operation on its inputs followed by a so-called activation function, which is always non-linear. We address the choice of layer-specific operations in section \ref{subsubsec:affine_trans} and then introduce the activation functions in section \ref{subsubsec:act_fun}. In general, the layer-specific operation is always the name-giving component for the layer, so for example if we compute a 2D convolution as the layer-specific operation on the input and then apply an activation function, we call the set of these two stages a \emph{convolutional layer}. Figure \ref{fig:nn_schematic} shows a neural network whose first layers are convolutional layers followed by a fully connected layer that produces the predictions.

\subsubsection{\label{subsubsec:affine_trans}Affine transformations}

All common choices for layer-specific operations are affine transformations. They all introduce trainable weights; free parameters that are adjustable during the training process and are sometimes called neurons due to the intuition that in a fully connected layer they share some similarity to the dendrites, soma, and axon of a biological neuron \cite{Arbib1987}. These trainable weights are the name-giving components in a \emph{neural} network.

Now, the goal of training a neural network is to optimize all these weights for all layers, so, that the predictions for all images in the training data match their accompanying original labels. The original labels are called ground truth and define the upper limit of how good a network can fit a domain. No neural network is better than its training data. In this section, we briefly illustrate the affine transformations of the fully connected layer and the convolutional layer and then explain in the next section the role of the activation function.

\paragraph{Fully connected layer}

The name-giving operation for the fully connected layer is a matrix multiplication performed on a flattened input, for example, a $m\times n$ sized input image would be \emph{flattened} into a $m \cdot n$ sized vector. Mathematically this is a matrix multiplication between a matrix and a vector:
\begin{gather}
 a_{j} = \sum_{k=1}^m x_{k}w_{kj},
 \label{eq:dense_layer}
\end{gather}
where $\bm{x}$ is the flattened input and $\bm{w}$ is the weight matrix of a fully connected layer. Here, all input vector elements (e.g., the pixels of an image, now arranged in one large row $x_{k}$) contribute to all output matrix elements and are therefore \emph{connected}. Furthermore, by convention is $x_{0}$ defined as \num{1} and $w_{0j}=b_{j}$, where $b_{j}$ is a free and trainable bias parameter.

\begin{figure*}
 \includegraphics[width=1\textwidth]{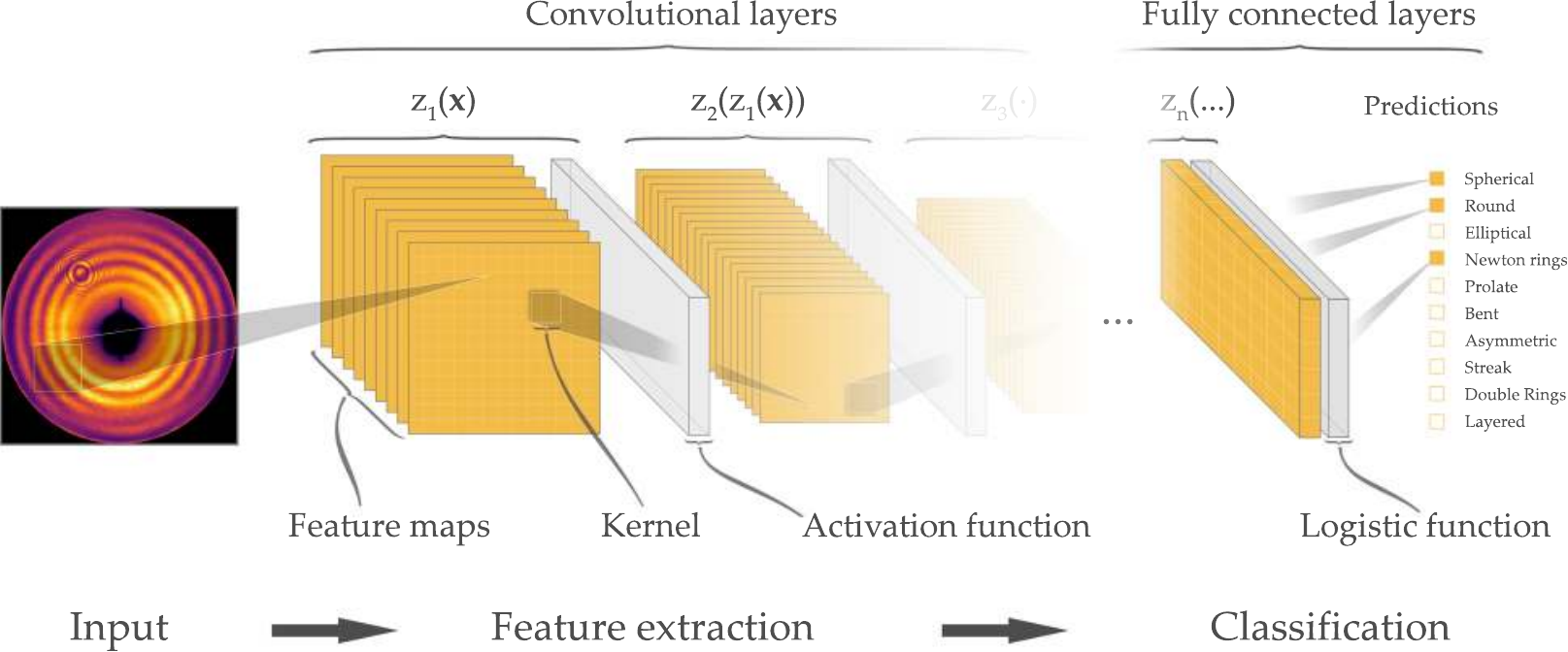}%
 \caption{\label{fig:nn_schematic} Schematic visualization of a convolutional neural network; It shows the hierarchical structure of the network with the function hierarchy $z_{1}, ~\dots ~z_{n}$ above each layer. Depicted as input is a diffraction image, which is getting expanded by \num{9} trainable convolutional kernel into \num{9} feature maps. Note, only \num{1} kernel, producing the last feature map, is shown. The output of the first layer is then passed through multiple convolutional layer, this is the feature extraction part of the neural network. Ultimately, a fully connected layer with a logistic function as activation function produces the predictions. Every layer consists of \num{2} stages, also indicated by the brackets underneath $z_{1}, ~\dots ~z_{n}$. The first stage is an affine transformation and the second one is a non-linear function, called an activation function. The operation that is used as affine transformation is then the name-giving component for the layer, e.g., a convolutional layer uses a convolution as affine transformation. The choice of the activation function is subject to empirical optimization with various choices possible. Section \ref{subsubsec:affine_trans} describes the affine transformations in more detail and section \ref{subsubsec:act_fun} covers the basics on activation functions.
 }
\end{figure*}
\paragraph{Convolutional layer}
In a convolutional layer, the trainable weights are parameters of a kernel that slides over the inputs, this is visualized in Figure \ref{fig:nn_schematic}.
The general idea of a convolutional layer is to preserve the spatial correlations in the input image when going to a lower dimensional representation (the next layer). This is achieved by using a kernel with a spatial extent larger than \num{1} px. The kernel size is then also the extent to which one kernel can correlate different areas of an input and is called its local receptive field.
Each kernel produces one output which is called a \emph{feature map} or \emph{filter}. Multiple feature maps from multiple kernels are grouped within one convolutional layer. For example, the first convolutional layer in Figure \ref{fig:nn_schematic} produces \num{9} feature maps out of the input diffraction image and hence has \num{9} kernels that get optimized during training. Since we usually only have in the input layer a 2-dimensional diffraction image as input and a high number of feature maps for every subsequent convolutional layer as their inputs, we define the output of a convolutional layer with a \num{4}-dimensional kernel $k$ that produces $i$ feature maps of size $j \times k$:
\begin{gather}
 a_{i,j,k}=\sum_{l,m,n}x_{l,j+m-1,k+n-1}k_{i,l,m,n},
 \label{eq:conv_filter}
\end{gather}
here the input $x$ has $l$ dimensions of size $j\times k$ and we slide a kernel of size $m \times n$ across all these $l$ dimensions. In the given example for the input layer, $l$ is simply \num{1} and the summation is just across one input image, as shown in Figure \ref{fig:nn_schematic}.

\subsubsection{\label{subsubsec:act_fun}Activation functions}
Regardless of the affine transformation that is used, all layer-specific operations produce trainable weights which are passed through an activation function. This function is always non-linear. We only address \num{2} activation functions here as they are the most common used by the community and the only ones we use; The sigmoid and the \emph{LeakyRelu} function. The first one is a logistic regression function used mostly at the outputs of neural networks, and the second one is a piecewise linear activation function used between layers for numerical reasons \cite{Nair2010,Maas2016}.
The sigmoid function is given as:
\begin{gather}
 h\left(x\right) = \frac{1}{1+\exp{\left(-x\right)}}
 \label{eq:sigmoid}
\end{gather}
and the \emph{LeakyRelu} function is given as:
\begin{gather}
 h(x) =
 \begin{cases}
  x        & \text{if $x\ge0$} \\
  \gamma x & \text{if $x<0$},
 \end{cases}
 \label{eq:LeakyReLu}
\end{gather}
where in both functions $x\in\bm{a}$ are the trainable weights of the affine transformation (the convolutional or the fully connected layer operation, i.e., the output of Equation \ref{eq:dense_layer} or \ref{eq:conv_filter}) and $\gamma$ is the slope for the negative part in the \emph{LeakyRelu} function and is called \emph{leakage}.

In Figure \ref{fig:nn_schematic} the last activation function of the neural network, denoted by \emph{Logistic function}, is a sigmoid function, because its output can be interpreted as a probability in a Bernoulli distribution, yielding a probability for how likely it is that a given event (an image in our case) is part of a class (in our case, the pre-defined classes from Table \ref{tab:helium_stats}). Sigmoid functions always give an output between \num{1} and \num{0}.
In our case, we have \num{11} distinct classes which are mutually non-exclusive, which means every image has a probability of being part of every class. Using a sigmoid function at the end of the neural network yields therefore 11 distinct Bernoulli distributions.
The generalization from the single-case Bernoulli distribution to its multi-case n-class distribution equivalent is called categorical distribution.

Interpreting the output of the neural network, as well as the original labels, as a categorical distribution is key to train the neural network because only then we can use statistical measures to evaluate the quality of the neural network's prediction, which allows us to optimize it iteratively.

However, due to the non-linearity of all activation functions, optimizing a neural network is a non-convex problem where no global extrema can be found with certainty. The general procedure is that of a forward pass and then a backward correction. Meaning, we feed the neural network several images, take the network's prediction and compare this prediction to the ground truth; This is the forward pass. Then we calculate a loss function which is a metric for how bad or good the predictions were, see the next section, and correct the weights of the network in a way that it would be better equipped to predict the labels for the images it just saw. This correction step is starting at the end of the network using an algorithm called backpropagation; hence the name backward correction, see section \ref{subsubsec:grad_des_back_prop}.

%
%

\subsubsection{\label{subsubsec:loss_function}The forward pass: Assess the network's predictions}
Optimizing a neural network always starts by feeding it multiple images and evaluate what the neural network made of it. For assessing the quality of the network's prediction a so-called loss function is used. It is the defining metric that we seek to minimize during the training of the neural network. In every training step, we compare the output of the neural network to the real labels provided by the researcher and calculate the so-called loss. Lower loss values correspond to a higher prediction quality of the neural net.

Therefore, the goal during the training process is to adjust all weights and biases within the network so, that the loss is minimal for all input training images. There are various possible loss functions which often serve a specific purpose. For classification tasks, such as the present case, primarily the \emph{cross-entropy} is used \cite{Schmidhuber2014,LeCun2015,Szegedy2016,Hinton2015}. Cross-entropy is a concept from information theory giving an estimate about the statistical distance between a \emph{true} distribution $p$ and an \emph{unnatural} distribution $q$. In our case, $p$ is the categorical distribution over the ground truth labels, and $q$ is the output of the neural network.


%

Cross-entropy is calculated as the sum of the Shannon entropy \cite{Shannon1948} for the \emph{true} distribution $p$ and the Kullback-Leibler divergence \cite{Kullback1951} between $p$ and $q$. The former is a measure of the total amount of information of $p$, and the latter is a typical distance measure between two probability distributions.

If the Kullback-Leibler divergence is zero, then the cross-entropy is just the Shannon entropy of $p$, and we have $p=q$. Then, the predictions of the neural network are not distinguishable from the labels of all training images.

Cross-entropy can be formally written as:
\begin{gather}
 H(p, q) = H(p) + D_{\mathrm{KL}}(p \| q)
 \label{eq:cross_entropy_first}
\end{gather}
where $H(p)$ is the Shannon entropy of $p$, and $D_{\mathrm{KL}}(p \| q)$ is
the Kullback–Leibler divergence of $p$ and $q$ \cite{Goodfellow2016a}.
%

When using a sigmoid function as activation function on the output layer, the final loss function can be defined as:
\begin{gather}
 H(x_{out}, x) = \sum_{i}^{M} x^{out}_{i}-x^{out}_{i}~x_{i}+\log{\left(1+\exp{\left(-x^{out}_{i}\right)}\right)},
 \label{eq:cross_entropy_binary}
\end{gather}
where $M$ is the number of all images in the training data, $x^{out}_{i}$ is the prediction for one image from the deep neural network and $x_{i}$ is the original label of the image, assigned by the researcher. Please see appendix \ref{app:derive_cross_entropy}  for a complete derivation.

Using Equation \ref{eq:cross_entropy_binary} as it is, would require us to pass all images through the network for one training step, as the sum runs over all images. This is computational intractable. Therefore, we use a variant of Equation \ref{eq:cross_entropy_binary} where the sum runs only over a stochastically chosen subset of size $bs$, called a batch. The size of that batch is called batch size and is an important hyperparameter that needs to be chosen prior to training, see section \ref{sec:training_routine}. One iteration step now involves only $bs$ images from the dataset, and we define an \emph{epoch} as the number of iteration steps it takes the network during the training to see all images one time.

To summarize, minimizing the cross-entropy is the goal during the training process in a neural network. The network learns to link the user-defined labels to the provided images. All that's left to understand the basic training process of a neural network is a way to adjust the weights in all layers.


\subsubsection{\label{subsubsec:grad_des_back_prop}The backward correction: Gradient descent and backpropagation}

Optimizing the weights within the neural network so that they give minimal loss for all training images is done using two distinct algorithms; gradient descent and backpropagation. In principal, gradient descent works by evaluating the gradient at some point and then moving a certain step-size in the opposite direction; This is done iteratively until the gradient is smaller than some pre-defined threshold, which is the numerical equivalent of calculating the extrema of a function analytically.

The basic gradient descent step is given by:
\begin{gather}
 \bm{w}_{\tau+1} = \bm{w}_{\tau}-\eta\nabla_{\bm{w}_{\tau}} H\left(x_{out}, x\right).
 \label{eq:gradient_descent}
\end{gather}
where $\eta$ is the afore mentioned step-size, called learning rate, $\nabla_{\bm{w}_{\tau}}$ is the gradient w.r.t. the weights at step $\tau$ and $H\left(x_{out}, x\right)$ is the loss function from Equation \ref{eq:cross_entropy_binary}.
With Equation \ref{eq:gradient_descent} we already could update the weights within the output layer of the neural network ($z_n(\cdot)$), since for the output layer we can calculate the numerical gradients. But we can't do this for the layers that come before the output layer since we're lacking a way to include these. In order to propagate the gradient descent correction throughout the network an algorithm called backpropagation is used \cite{Rumelhart1986}:

First, we define the gradient of $H(x_{out}, x)$, w.r.t. the weights at the output of the deep neural network, using the chain rule:
\begin{gather}
 \nabla_{\bm{w}_{\tau}} H\left(x_{out}, x\right) = \frac{\partial H(x_{out}, x)}{\partial w_{j\tau}^{N}} = \frac{\partial H(x_{out}, x)}{\partial h^{N} \left( a_{j}^{N}\right)} \frac{\partial h^{N} \left( a_{j}^{N}\right)}{\partial w_{j\tau}^{N}},
 \label{eq:app_initial_partial}
\end{gather}
where $N$ denotes the layer depth of the output layer, $h^{N}(\cdot)$ is the used activation function in that layer and $a^{N}_{j}$ are the outputs of the layer-specific operation, as in Equation \ref{eq:dense_layer} and \ref{eq:conv_filter}.
Starting from there we include the layer, preceding the output layer ($z_{n-1}(z_n(\cdot))$), by making use of the
chain rule again:
\begin{gather}
 \frac{\partial H(x_{out}, x)}{\partial h^{N} \left( a_{j}^{N} \right)} = \frac{\partial H(x_{out}, x)}{\partial h^{N-1} \left( a_{j}^{N-1} \right)} \frac{\partial h^{N-1} \left( a_{j}^{N-1} \right)}{\partial h^{N} \left( a_{j}^{N} \right)}.
 \label{eq:app_additional_partial}
\end{gather}
This can be iteratively repeated until the input layer ($z_{1}(\cdot)$) is included in the calculation.
By making use of the chain rule until we reach the input layer we can include all trainable weights of all layers into the correction term of the gradient descent algorithm. With this, we conclude the full optimization routine in Table \ref{tab:iterative_optimization_routine}.

\begin{table}[tbh]
 \caption{\label{tab:iterative_optimization_routine}
  The iterative optimization routine for a deep feed-forward neural network.}
 \begin{ruledtabular}
  \begin{tabular}{ll}
   \num{1}. & \textbf{Forward pass}: Propagate $bs$ images through the network.       \\
   \num{2}. & \textbf{Evaluate the predictions}: At the output layer calculate        \\
   ~        & the loss between the ground truth and the output of the                 \\
   ~        & deep neural network (Equation \ref{eq:cross_entropy_binary}).           \\
   \num{3}. & \textbf{Construct the backpropagation rule}: Include all                \\
   ~        & gradients w.r.t. the weights of all layer according                     \\
   ~        & to Equation \ref{eq:app_additional_partial}.                            \\
   \num{4}. & \textbf{Backward correction}: Update all weights in the                 \\
   ~        & network using gradient descent, see Equation \ref{eq:gradient_descent}.
  \end{tabular}
 \end{ruledtabular}
\end{table}

\subsubsection{\label{sec:training_routine}Training setup}

Of significant importance is the way how the network is constructed; How deep should the network be and of what should it consist? For nomenclature, the combination of all used layers, the depth of the network and the used activation functions is called an architecture.

We benchmarked the performance of various architectural choices when used with diffraction images as input and provide the results in appendix \ref{subsec:architecture_choices} and not in the main paper, due to its rather technical character. In short, all architectures are established through extensive empirical research. So far, not only the leading A.I. research institutes, like the Massachusetts Institute of Technology (MIT) or the University of Toronto, but also large companies like Google, Facebook and Microsoft have invested significant amounts of resources to establish well working \emph{out-of-the-box} solutions \cite{Szegedy2016,Schmidhuber2014,LeCun2015}.

Building on this and after extensively benchmarking the most common architectures on our own, we settled on an architecture called pre-activated wide residual convolutional neural network in its \num{18}-layer configuration, called \emph{ResNet18} \cite{He2015, He2016a, Zagoruyko2016}. In essence, it is a convolutional neural network much like the example in Figure \ref{fig:nn_schematic} but it employs so-called residual skip connections which increase \emph{Accuracy} while decrease training time, see appendix \ref{subsec:architecture_choices} for further details as well as comparisons with other architectures.

After settling on an architecture, training a neural network requires fine-tuning of multiple free parameters. Four of them are critical: The learning rate $\eta$, the batch size \emph{bs} and so-called regularization parameters of which we have two (which will be introduced at the end of this section).

We set the initial learning rate for the gradient descent algorithm to $\eta = 0.1$, see also Equation \ref{eq:gradient_descent}. Throughout the training we multiply $\eta$ with \num{0.1} every \num{50} epochs, this increases the chance for the gradient descent algorithm to get numerically closer to a minimum in the loss function \cite{He2015}. Furthermore, we use a batch size of \num{48} for all training procedures, see also the explanations for Equation \ref{eq:cross_entropy_binary}.

We split the manually classified part of the helium dataset into a training and an evaluation subset, where we shuffle the order of all images and then select \SI{85}{\percent} for the training set while the rest serves as an evaluation set.

We rescale all diffraction images to \SI{224x224}{\pixel} which is necessary to fit the deep neural net on two Nvidia \num{1080}Ti GPUs, each having \num{11} GB memory. The image dimensions are chosen to be a compromise between file size and resolution. All features we are training the neural network on are still clearly visible and distinguishable after the rescaling.


Furthermore, we face the problem of having a comparatively small training set, consisting of only $\approx 6000$ classified images, which could result in a phenomenon called over-fitting. Meaning the network memorizes the training set without learning to make any meaningful prediction from it. Therefore we employ two additional techniques called regularization and data augmentation:
\begin{enumerate}
 \item Regularization means adding a so-called penalty term to the loss function. There are two regularizations we use, L1 and L2 \cite{Zou2005}. These penalty terms are dependent on the weights themselves and not on the labels, making the loss function explicitly dependent on the weights of the neural network. This dependency encourages the neural network to reduce the values of all weights according to the two penalty terms and ultimately find a sparser solution which in return helps to prevent over-fitting. Formally we add these two terms to the loss in Equation \ref{eq:cross_entropy_binary}:
       \begin{gather}
        H(x_{out}, x)_{reg} = H(x_{out}, x) + \alpha ||\bm{w}||_{1}  + \beta ||\bm{w}||_{2}
       \end{gather}
       where $H(x_{out}, x)$ is the cross-entropy loss function, $||\bm{w}||_{1}$ and $||\bm{w}||_{2}$ are the L1- and the L2-norm applied on the sum of all trainable weight parameters and $\alpha$ and $\beta$ are so-called regularization coefficients. In our experiments we set $\alpha$ and $\beta$ to \num{1e-5} during training. Using L1 and L2 regularization in combination is commonly referred to as elastic net regularization \cite{Zou2005}.
 \item Data augmentation means creating artificial input images by randomly applying image transformations on the original image like flipping the vertical or the horizontal axes and adjusting contrast or brightness values randomly. This greatly increases the robustness to over-fitting and is used as a standard procedure when facing small training datasets \cite{Hinton2012,Perez2017}.
\end{enumerate}

We were able to train deep neural networks with a depth of up to \num{101} layers without over-fitting using regularization and data augmentation, see appendix \ref{subsec:architecture_choices}. In all experiments reported here we choose a depth of \num{18} layers for the neural network, due to numerical, memory and time reasons.
We trained all deep neural networks variants for \num{200} epochs.

\subsection{\label{subsec:evaluating_neural_network}Evaluating a deep neural network}
We use three metrics to assess the quality of the predictions from the neural network, \emph{Accuracy}, \emph{Precision}, and \emph{Recall}. We calculated these metrics every \num{2500} training iteration steps ($\approx 52$ epochs) using the evaluation dataset. \emph{Accuracy} is formally defined as:

\begin{gather*}
 \text{Accuracy} = \frac{\text{True Positives} + \text{True Negatives}}{\text{Condition Positives} + \text{Condition Negatives}},
\end{gather*}

where condition positives/negatives is the real number of positives/negatives in the
data and true positives/negatives is the correct overlap of the prediction from the model and the condition positives/negatives.
An \emph{Accuracy} of \num{1} corresponds to a model that was able to predict all classes of all images correct. Therefore, \emph{Accuracy} is a good measure for evaluating the prediction capabilities of a model when true positives \emph{and} true negatives are of importance. Predicting negative labels correct is in the case of the helium dataset of particular interest because we want to estimate if the neural network was able to understand the complex inter-class relationships imposed by the researcher. The network should realize that if, for example, one prediction is \emph{Spherical/Oblate}, it cannot simultaneously be \emph{Prolate}. Therefore, the network has to produce a true negative for either one of these predictions. However, using only \emph{Accuracy} as a metric has several downsides. The most important one is the decreased expressiveness of \emph{Accuracy} when working in a multi-class scenario. In order to understand this, we first introduce \emph{Precision} and \emph{Recall}, and then provide an example:
\begin{gather*}
 \text{Precision} = \frac{\text{True Positives}}{\text{True Positives}+\text{False Positives}},\\[1em]
 \text{Recall} = \frac{\text{True Positives}}{\text{True Positives}+\text{False Negatives}}.
\end{gather*}
\emph{Precision}, also-called positive predictive value, is a measure for how reasonable the estimates of the model were when it labeled a class positive, and \emph{Recall} is a measure for how complete the model's positive estimates were.

For example, if the model would predict all training images in the helium dataset to be \emph{Spherical/Oblate} and nothing else (out of \num{7264} images, \num{6589} are indeed \emph{Spherical/Oblate}) then \emph{Accuracy} would be \num{0.767}, which translates to \SI{77}{\percent} of all labels correctly assigned. However, if the model estimated all images to be part of no class (setting every label to negative), then \emph{Accuracy} would be \num{0.801}, because out of \num{79904} possible labels (\num{11} independent classes for \num{7264} images), \num{64339} are negative. Therefore, we would have a useless model that still was able to predict \SI{80}{\percent} of all labels correct.

Using \emph{Precision}  in these both examples would give \num{0.907} for the \emph{Spherical/Oblate} example and \num{0.000} for the all-negative example. \emph{Precision} is, therefore, a metric that quantifies how well the positive predictions were assigned. Since \SI{91}{\percent} of all images are indeed \emph{Spherical/Oblate}, setting all labels positive in the \emph{Spherical/Oblate} class can make sense, and \emph{Precision}  also provides insight when the model makes no positive prediction at all which would be a useless model for our purpose. However, \emph{Precision}  alone is not sufficient as a metric. At this point we don’t know if our model predicted almost every possible positive label correct or if only a small fraction of all positive labels were assigned correctly, we, therefore, need an additional measure for the generalization capabilities of our model. For that reason, \emph{Precision}  is always used in combination with \emph{Recall}. The \emph{Recall} for our first example is \num{0.423} and for the second one \num{0.000}. \emph{Recall} relies on False Negatives instead of the False Positives, used by \emph{Precision} , which provides a measure about the completeness of all positive predictions compared to all positive labels within our data. \emph{Recall} states that our model only captured \SI{42}{\percent} of all possible positive labels in the \emph{Spherical/Oblate} example, showing that generalization of the model would not be sufficient for a real-world application.

Therefore, a balanced interpretation of these three metrics is necessary to estimate the quality of the models tested here.

\section{\label{sec:deep neural network_baseline}Baseline performance of neural networks with CDI data}

In this section we briefly report on what we call baseline results. We used the previously described ResNet \cite{He2016a} neural network architecture in its basic configuration with a depth of 18 layers, termed vanilla configuration or ResNet18 (see section \ref{sec:training_routine}) and trained it with the helium diffraction data set as described in section \ref{sec:dataset} as well as with a reference data set from the literature \cite{Kassemeyer2012}.
This reference data set was made freely available on the CXIDB by \citet{Kassemeyer2012} \footnote{\url{https://www.cxidb.org/id-10.html}}. It contains diffraction patterns of a number of prototypical diffraction imaging targets, namely the \emph{Paramecium bursarium Chlorella virus} (PBCV-1), \emph{bacteriophage T4}, \emph{magnetosomes} and \emph{nanorice}. For further experimental details see \citet{Kassemeyer2012}.

We selected this dataset because of a previous publication dealing with this dataset \cite{Bobkov2015}, that describes, to our knowledge, the current state-of-the-art method for classification and sorting of diffraction images \cite{Bobkov2015}. \citet{Bobkov2015} trained a support-vector-machine on the CXIDB dataset and inferred the particle type directly from the diffraction images. Overall, they achieved an \emph{Accuracy} of up to \num{0.87}, but only on selected high quality images with a high confidence score of the support-vector-machine above \num{0.75}.




Table \ref{precision_recall_overall_metrics_architectures} shows the overall evaluation metrics as well as the training wall time. $\text{Train Time}_{Max}$ is the time when the neural network achieved the highest \emph{Accuracy} score on the evaluation dataset, and $\text{Train Time}_{Full}$ is the time for training 200 epochs.
In practice, we achieved optimal convergence after training for 70 to 100 epochs.


We achieved an \emph{Accuracy} of \num{0.967} on not only a high quality subset of the CXIDB data, like in \cite{Bobkov2015}, but on all available data (see table
\ref{precision_recall_overall_metrics_architectures}), using a vanilla ResNet18 architecture, proving that using a neural network significantly
outperforms the current state-of-the-art approach in \cite{Bobkov2015}.

\begin{table}[tbh]
 \caption{\label{precision_recall_overall_metrics_architectures}
  Overall evaluation metrics for the ResNet18 architecture (vanilla configuration) and both datasets. The table gives the max values during training for \emph{Accuracy}, \emph{Precision} , and \emph{Recall}. The training time after which the neural network achieved the highest \emph{Accuracy} score on the evaluation dataset is labeled $\text{Train Time}_{Max}$ and the time for training the full \num{200} epochs is labeled $\text{Train Time}_{Full}$. See also appendix \ref{subsec:architecture_choices} for further details.}
 \begin{ruledtabular}
  \begin{tabular}{rcc}
   Architecture                   & \multicolumn{2}{c}{ResNet18}               \\
   Dataset                        & CXIDB                        & Helium      \\
   \colrule
   \multicolumn{3}{c}{}                                                        \\
   \emph{Accuracy}                & \num{0.967}                  & \num{0.955} \\
   \emph{Precision}               & \num{0.932}                  & \num{0.918} \\
   \emph{Recall}                  & \num{0.933}                  & \num{0.866} \\
   $\text{Train Time}_{Max}$ [h]  & \num{0.278}                  & \num{0.231} \\
   $\text{Train Time}_{Full}$ [h] & \num{0.668}                  & \num{0.694}
  \end{tabular}
 \end{ruledtabular}
\end{table}

In the case of the helium dataset we face a much more complicated multi-class learning problem (one image can belong to multiple classes compared to one image belongs to exactly one class as it is in the CXIDB data). However, we reach a comparable \emph{Accuracy} score of \num{0.955}. Even more promising, \emph{Precision}  and \emph{Recall} are very high for the helium and the CXIDB dataset, proving that the neural network not only predicted the true positives with high confidence and reliability (high \emph{Precision} ), it did so for almost all true positive labels in the evaluation dataset (high \emph{Recall}).

In the next section we show how to further improve on the baseline performance of neural networks with diffraction images as input data.

\section{\label{sec:deep neural network_for_scattering}Adapting neural networks for CDI data}

Here, we describe our contribution for using neural networks in combination with diffraction images.

First, we show in section \ref{subsec:activation_function} that the performance of a neural network can be enhanced when using a special activation function after the input layer.

Second, in section \ref{subsec:size_of_training_set} we benchmark the performance of the neural network when using a smaller amount of training data. The idea is to provide an intuition about how much the prediction capabilities deteriorate when a smaller training dataset is used. This is useful because so far a researcher still has to invest a lot of time preparing the training dataset and, more general, minimizing the time spent looking through the raw data is the ultimate goal for using a neural network in the first place.

Third, in section \ref{subsec:ccf} we propose a novel data augmentation in the form of a custom two-point cross correlation map that hardens the network against very noisy data. We show that when using this augmentation the network is more robust to noise from a uniform distribution added on top of the original diffraction image. This simulates the experimental scenario in which a very low signal-to-noise ratio is unavoidable, e.g., during CDI experiments with very limited photon flux \cite{Rupp2017} or very small scattering cross sections as it is the case with upcoming CDI experiments on single biomolecules \cite{Ikeda2012, Shintake2008}.


\subsection{\label{subsec:activation_function}The logarithmic activation function}

One of the key additions of this paper is the proposed activation function, formally stated in Equation \ref{eq:log_activation_function}. It is designed to account for the inherent property of diffraction images of scaling exponentially. More general, the intensity distribution of scattered light on a flat detector follows two laws, depending on the scattering angle that is recorded. For very small angles (SAXS and USAXS experiments) the Guinier approximation is the dominant contribution to the recorded intensity, while for larger scattering angles (SAXS and WAXS experiments) Porod's law becomes dominant \cite{Hammouda2010,Sinha1988}. Where the scattering intensity in the Guinier approximation is proportional to $\approx \exp{\left( -q^{2} \right)}$, in Porod's law the intensity scales with $\approx q^{-d}$. $q$ is the scattering vector (function of the scattering angle and of the wavelength in use) and $d$ is the so-called Porod coefficient, which can vary significantly depending on the object from which the light was scattered \cite{Hammouda2010}.

In any case, the recorded detector intensity for diffraction images scale exponentially. For this reason we propose a logarithmic activation function of the form:
\begin{gather}
 h(x) =
 \begin{cases}
  \alpha \left(\log{\left(x+c_{0}\right)}+c_{1}\right)  & \text{if $x\ge0$} \\
  -\alpha \left(\log{\left(c_{0}-x\right)}+c_{1}\right) & \text{if $x<0$},
 \end{cases}
 \label{eq:log_activation_function}
\end{gather}
where $\alpha>0$ is a tunable scaling parameter, $c_{0}=\exp{\left(-1\right)}$, $c_{1}=1$  and $x$ is the input.

We define $c_{0}$ and $c_{1}$ so that the activation function is anti-symmetric around 0, which helps speed up training and avoids a bias shift for succeeding layers \cite{Clevert2015, Szegedy2014}.



Since we are using a gradient-based optimization technique we need to take care that the gradient can propagate throughout the whole network, otherwise it would lead to so-called \emph{gradient flow} problems, which befalls deep architectures \cite{Hochreiter2001, Nair2010}. There are two possibilities for insufficient gradient flow, either the gradients are getting too small (\emph{vanishing gradient}) or too large (\emph{exploding gradient}) when propagating throughout the network.
Both scenarios lead to numerical instabilities during training making convergence for large architectures very hard or even impossible. The reason for this is the backpropagation algorithm which invokes the chain rule for calculating the gradients. Every gradient is therefore also a multiplicative factor for the gradient of a succeeding layer.
For our case the derivative of Equation \ref{eq:log_activation_function} w.r.t. $x$ is given by:
\begin{gather}
 \frac{\partial h(x)}{\partial x} =
 \begin{cases}
  \frac{\alpha}{x+c_{0}} & \text{if $x\ge0$} \\
  \frac{\alpha}{c_{0}-x} & \text{if $x<0$}.
 \end{cases}
 \label{eq:partial_log_activation_function}
\end{gather}
It shows that the gradient scales with $x^{-1}$ with a discontinuity of
size $\alpha~c_{0}^{-1}$ at \num{0}.

If we used this activation function for all activations throughout the network, the gradient would have an increased probability to vanish - or explode - the deeper the architecture gets.
In addition to that, the discontinuity at $x=0$ could lead to gradient jumps, which would further decrease numerical stability.
Therefore, we use the logarithmic activation function only for the first convolutional layer and use a \emph{LeakyRelu} activation with \emph{leakage} of \num{0.2} on all hidden layers.
This compromise still captures the exponential scale of the diffraction images but without losing numerical stability.

Since $\alpha$ is a tunable hyperparameter, we conduct experiments with three values for $\alpha \in \left[0.2, 0.5, 1.0\right]$ and evaluate its impact on the performance of the neural network.

In Table \ref{precision_recall_overall_activation_function} we provide the evaluation metrics for ResNet18 used with the logarithmic activation function, trained with three different values for $\alpha$. For comparison, we also provide the results of the unmodified ResNet18 labeled \emph{unmodified}. The best performing configuration is with an $\alpha$ value of \num{0.2}, maxing out with an \emph{Accuracy} of \num{0.965}. Therefore, providing a boost in \emph{Accuracy} of a full percentage point compared to the unmodified ResNet18. The lowest value for the maximum \emph{Accuracy} was reached without the logarithmic activation function, topping at \num{0.955}. \emph{Precision} and \emph{Recall} both increase with the addition of the logarithmic activation function. These improvements all come without increasing training time or complexity of the model.
\begin{table}[tbh]
 \caption{\label{precision_recall_overall_activation_function} Evaluation metrics for the ResNet18 network with and without the logarithmic activation function. We benchmark three values for $\alpha$. Results are shown for both datasets and are the maximum value recorded during training. Bold numbers indicate the best scores across their respective category.}
 \begin{ruledtabular}
  \begin{tabular}{rcccc}
   Architecture     & \multicolumn{4}{c}{ResNet18}                                                   \\
   $\alpha$         & \num{0.2}                    & \num{0.5}    & \num{1.0}    & \emph{unmodified} \\
   \colrule
   \multicolumn{5}{c}{}                                                                              \\
   Dataset          & \multicolumn{4}{c}{Helium}                                                     \\
   \emph{Accuracy}  & $\bm{0.965}$                 & \num{0.960}  & \num{0.959}  & \num{0.955}       \\
   \emph{Precision} & $\bm{0.922}$                 & \num{0.920}  & $\bm{0.922}$ & \num{0.918}       \\
   \emph{Recall}    & $\bm{0.870}$                 & $\bm{0.870}$ & \num{0.868}  & \num{0.867}       \\
  \end{tabular}
 \end{ruledtabular}
\end{table}
The maximum achieved \emph{Accuracy} seems to be anti-correlated to $\alpha$, with the $\text{ResNet18}_{\alpha=1.0}$ variant performing worst. We suspect that this is related to the smaller size of the discontinuity of the derivative of $h(a_{j})$ when choosing a small value for $\alpha$, see Equation \ref{eq:partial_log_activation_function}.

However, choosing even smaller values for $\alpha$ did not improve the \emph{Accuracy} further, either because the benefit from the activation function plateaus there or because we reached the classification capacity of this ResNet layout.


These results show convincingly that the addition of the logarithmic activation function improves the overall performance and generalization of the deep neural network. This is in so far expected because we imposed a form of feature engineering on the network, by exploiting a known characteristic of the dataset. Therefore, without increasing the complexity, the depth or the training time, we showed that using the logarithmic activation improves all relevant evaluation metrics. For this reason, we use the logarithmic activation function with an $\alpha$ value of \num{0.2} as default for all following experiments.

\subsection{\label{subsec:size_of_training_set}Size of the training set}



In this section, we evaluate the impact of the training set size on the evaluation metrics, we trained the $\text{ResNet18}_{\alpha=0.2}$ with a varying amount of labeled images. The reason for this is to provide intuition for how many images are needed to be classified manually before the employment of a neural network is useful. We uniformly select images from the training set but kept the same evaluation dataset described in section \ref{sec:training_routine}. We decreased the size of the training set in three stages (to \SI{75}{\percent} $\equiv$ \num{4631}, to \SI{50}{\percent} $\equiv$ \num{3088} images, to \SI{25}{\percent} $\equiv$ \num{1544} images).

\begin{table}[b]
 \caption{\label{precision_recall_overall_training_set_size} Evaluation metrics of the $\text{ResNet18}_{\alpha=0.2}$ network with the logarithmic activation function and an $\alpha$ value of \num{0.2}. Results are shown for the helium datasets and reflect the maximum achieved value reached throughout the training process, assessed on the evaluation dataset. Bold numbers indicate the best scores across their respective category.}
 \begin{ruledtabular}
  \begin{tabular}{rcccc}
   Architecture      & \multicolumn{4}{c}{$\text{ResNet18}_{\alpha=0.2}$}                                           \\
   Training set size & \num{6174}                                         & \num{4631}  & \num{3088}  & \num{1544}  \\
   \colrule
   \multicolumn{5}{c}{}                                                                                             \\
   Dataset           & \multicolumn{4}{c}{Helium}                                                                   \\
   \emph{Accuracy}   & $\bm{0.965}$                                       & \num{0.915} & \num{0.829} & \num{0.797} \\
   \emph{Precision}  & $\bm{0.922}$                                       & \num{0.821} & \num{0.740} & \num{0.673} \\
   \emph{Recall}     & $\bm{0.870}$                                       & \num{0.771} & \num{0.679} & \num{0.593} \\
  \end{tabular}
 \end{ruledtabular}
\end{table}

Table \ref{precision_recall_overall_training_set_size} shows the performance of $\text{ResNet18}_{\alpha=0.2}$ when trained with datasets of different sizes. For the helium dataset, the maximum achieved \emph{Accuracy} is dropping from \numrange{0.965}{0.797} when using only \num{1544} images instead of the full \num{6174} images. Even more pronounced is the decline in \emph{Precision}  and \emph{Recall} from \num{0.922} and \num{0.870} to \num{0.673} and \num{0.593} for the smallest training set size.
The steeper decline rate for \emph{Precision}  and \emph{Recall}, compared to \emph{Accuracy}, can be understood as the helium dataset predominantly consists of \emph{Negative} ground truth labels (\num{64 339} out of \num{79 904} labels) to which the neural networks resorts in the absence of sufficient training data. \emph{Precision} and \emph{Recall}, on the other hand, provide only information about the positive prediction capabilities and their completeness and therefore decrease faster when a smaller training set size is used.

This shows that the number of images is critical for the prediction capabilities of the neural network.
The drastic decrease in training set size results in a much worse generalization of the model, detecting only those images that are very close to the ones from the training set, missing most from the evaluation set. The network has not learned the characteristics of a particular class to a point where it can transfer the gained knowledge to other images, which is the one critical property for which we employed a neural network in the first place.

Therefore, if time is limited, one may be well advised to concentrate efforts on preparing a sufficiently large, high-quality training dataset while using e.g. our here presented neural network approach in its standard configuration.


\subsection{\label{subsec:ccf}Using two-point cross-correlation maps to be more robust to noise}

This section introduces an image augmentation based on the two-point cross-correlation function, which increases the resistance to noise. We prepare four training sets, each with an increasing amount of noise sampled from a uniform distribution and analyze the noise dependence of the neural network.


One of the principal problems in CDI experiments, or imaging experiments in general, is recorded noise. Noise often leads to computational problems due to noise resistance being a known weak point for a significant fraction of predictive algorithms \cite{Atla2011}. In particular, deep neural networks are known to be easily fooled by noise. When adding noise to an image, whose addition may be invisible to the human eye, a neural network can come to entirely different conclusions and this even with high confidence; Seeing a panda where there was a wolf \cite{Nguyen2014,Moosavi-Dezfooli2016}. Therefore, we propose an additional pre-processing step for the input images to increase the noise resistance of the neural network.

\begin{figure*}
 \includegraphics[width=1\textwidth]{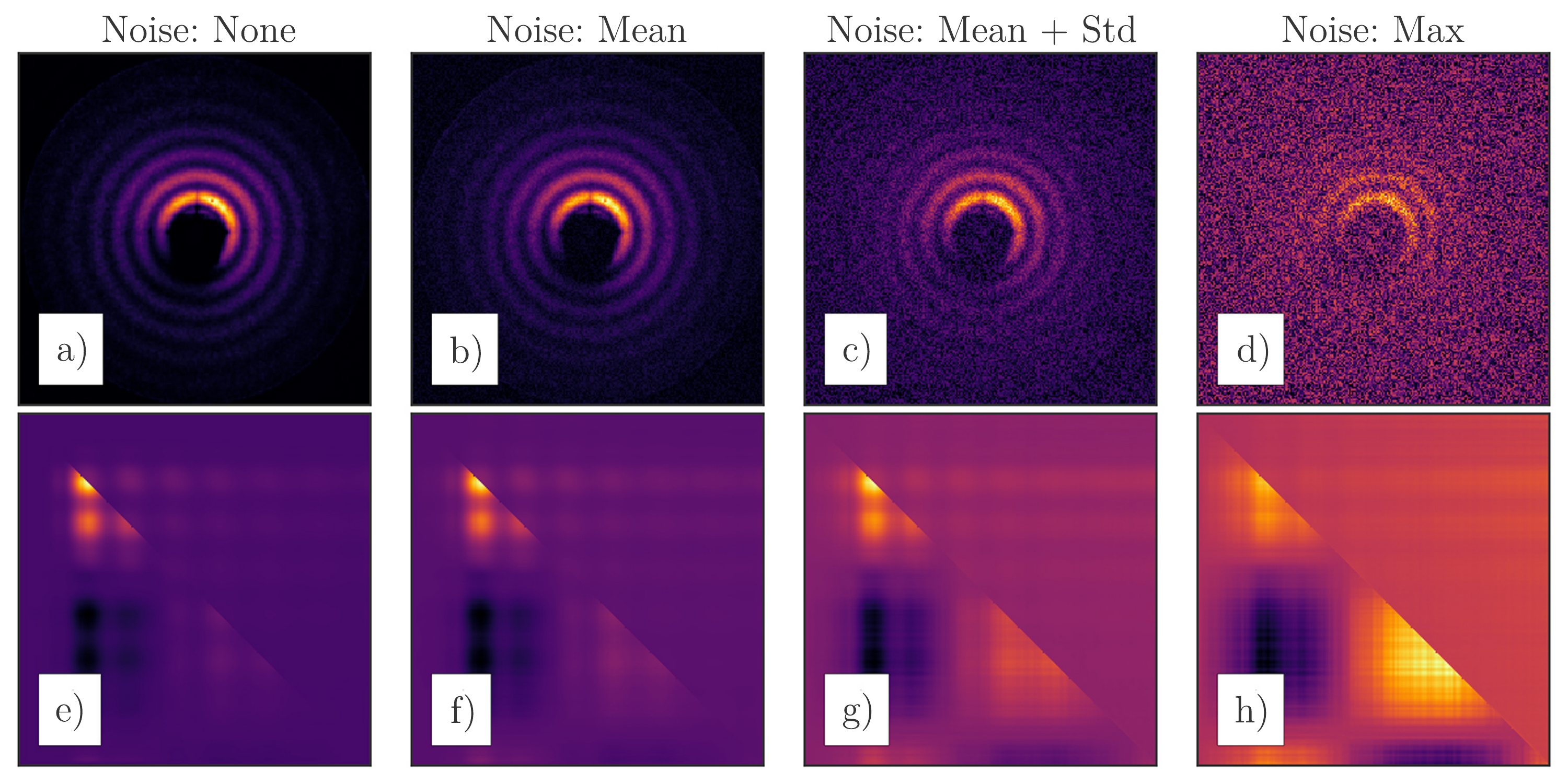}
 \caption{\label{fig:noise_plot} a) to d) showing the various stages of added noise to a standard scattering image. e) to h) are the calculated correlation maps with the upper triangle of order $n=8$ and lower triangle from the full CCF calculation.}
\end{figure*}

To quantify the quality of an image, the signal-to-noise ratio is often used. It is a measure for how much noise is present when compared to some information content, where low values indicate that information might be indistinguishable from noise. It has been shown that higher orders of the two-point cross-correlation function (CCF) can act as a frequency dependent noise filter and increase the quality of a reconstruction of a diffraction image even in the presence of recorded noise\cite{Kurta2017,Donatelli2015}. And since the CCF can be interpreted as an image, see Figure \ref{fig:noise_plot} e) to h), we employ this method in a similar manner to optimize the use-case with a convolutional deep neural network, expecting that the higher-order terms make the neural network more resistant to the presence of noise.

In general, the CCF is defined as:
\begin{gather}
 C_{i, j}\left(q_{i} , q_{j} , \Delta\right) = \int_{-\infty}^{\infty} I_{i}^{*}\left(q_{i} , \phi\right) I_{j}\left(q_{j} , \phi + \Delta\right) d\phi ,
 \label{eq:ccf}
\end{gather}

where $\Delta$ is the angular separation, $\phi$ is the angular coordinate, and $\left(i, j\right)$ denotes the index of the two scattering vectors $q_{i}$ and $q_{j}$. For discrete $\phi$ and written as Fourier decomposition, Equation \ref{eq:ccf} yields \cite{Kurta2017}:

\begin{gather}
 C_{i, j}^{n}\left(q_{i},q_{j}\right) = I_{i}^{n*}\left(q_{i}\right)~I_{j}^{n}\left(q_{j}\right),
 \label{eq:ccf_fc}
\end{gather}
where $n$ denotes the order of the CCF. $I_{i}^{n}$ is given by:
\begin{gather}
 I_{i}^{n}\left(q_{i}\right) = \frac{1}{2\pi}\int^{2\pi}_{0}I\left(q_{i},\phi\right)\exp\left(-in\phi\right)
 d\phi
 \label{eq:ccf_intensity}
\end{gather}

Since $C_{i, j}=C_{j, i}$, we can split the final correlation map into an upper and a lower triangle matrix. To maximize information, and to optimally use the local receptive fields of the convolutional layers, we merge the lower triangle from the full CCF calculation, Equation \ref{eq:ccf} with $\Delta=0$, and the upper triangle of order $n=8$ from Equation \ref{eq:ccf_fc}. Therefore, we combine a \emph{plain} correlation map with a higher order map that is more resistant to noise, see Figure \ref{fig:noise_plot} e) to h) for a full example.

To test the robustness of this method, we use the $\text{ResNet18}_{\alpha=0.2}$ and train it with various pre-processed datasets.

From our original dataset we derive three additional datasets that only differ in the amount of noise added. We do this as follows; First, we calculate the mean, the standard deviation (std) and the maximum intensity values of each image in the original dataset. From these values we calculate the median, instead of the mean (due to increased robustness against outliers); ending up with three statistical characteristics describing the intensity distribution throughout all diffraction images.
With that, we define three continuous uniform distributions to sample noise from. A continuous uniform distribution is fully defined by an upper and a lower boundary; $a$ and $b$, respectively. The probability for a value to be drawn within these boundaries is equal and non-zero everywhere. For our three noise distributions we always use a lower boundary of \num{0} and vary the upper boundary so that $b$ is either \emph{the mean}, \emph{the mean + the std.} or \emph{the maximum} of the intensity distribution on the images (the three statistical characteristics described above).

For example, for creating \emph{the maximum} noise dataset, we looped through every diffraction image and added noise sampled from \emph{the maximum} noise distribution. We do this for all three noise distributions. From these three noise embedded datasets, as well as our original dataset, we calculate the here proposed CCF maps. This leads to a total of eight data sets; for each of them we train a $\text{ResNet18}_{\alpha=0.2}$. An example of one image in all eight datasets is in Figure \ref{fig:noise_plot}.


%
%

\begin{table*}[hbt!]
 \caption{\label{precision_recall_ccf_overall} Evaluation results when training a $\text{ResNet18}_{\alpha=0.2}$ on the original diffraction images and on CCF maps calculated from them. The results reflect the maximum value achieved throughout the training process, assessed on the evaluation dataset. Bold numbers indicate the best scores across their respective category.}
 \begin{ruledtabular}
  \begin{tabular}{rcc|cc|cc|cc}
   Architecture     & \multicolumn{8}{c}{$\text{ResNet18}_{\alpha=0.2}$}                                                                                                                                                   \\
   Noise added      & \multicolumn{2}{c}{None}                           & \multicolumn{2}{c}{Mean} & \multicolumn{2}{c}{Mean + Std.} & \multicolumn{2}{c}{Max.}                                                           \\
   Input data       & CCF Maps                                           & Diff. Imgs.              & CCF Maps                        & Diff. Imgs.              & CCF Maps     & Diff. Imgs. & CCF Maps     & Diff. Imgs. \\
   \colrule
   \multicolumn{9}{c}{}                                                                                                                                                                                                    \\
   Dataset          & \multicolumn{8}{c}{Helium}                                                                                                                                                                           \\
   \emph{Accuracy}  & \num{0.950}                                        & $\bf{0.965}$             & \num{0.948}                     & $\bf{0.954}$             & $\bm{0.946}$ & \num{0.944} & $\bf{0.944}$ & \num{0.926} \\
   \emph{Precision} & \num{0.901}                                        & $\bf{0.922}$             & \num{0.897}                     & $\bf{0.910}$             & $\bm{0.893}$ & \num{0.887} & $\bf{0.905}$ & \num{0.865} \\
   \emph{Recall}    & \num{0.838}                                        & $\bf{0.870}$             & \num{0.833}                     & $\bf{0.853}$             & $\bm{0.823}$ & \num{0.815} & $\bf{0.814}$ & \num{0.808} \\
  \end{tabular}
 \end{ruledtabular}
\end{table*}

The results for these eight data sets are given in Table \ref{precision_recall_ccf_overall}. The performance of the neural network without added noise is much stronger when using the original diffraction images instead of the CCF maps. However, as soon as noise is added, the performance of the neural network trained on diffraction images deteriorates much faster as compared to the performance with CCF maps as input. When the upper boundary of the added noise excels the median values of \emph{mean + std.}, the neural network is performing better with the CCF maps instead of the original diffraction images. Especially with the noisiest dataset the differences in performance are significant. \emph{Precision} is increased by \num{4} percentage points when using the CCF maps as input, showing that our data augmentation may serve as a helpful asset when dealing with very noisy data.

In general, it is a viable alternative to use the CCF maps as input to the convolutional deep neural network, which should be considered an option in the case of very noisy data where it provides a boost to classification results. The downside is, calculating the CCF for every image comes at an additional computational cost. It took us three full days to calculate the CCF maps for all \num{39 879} images of both datasets on an Intel 6700K quad-core machine using a multi-threaded Python script (Also released on Github).

\section{\label{sec:interpretation_heatmaps}What the neural network saw}

\begin{figure*}
 \includegraphics[width=1\textwidth]{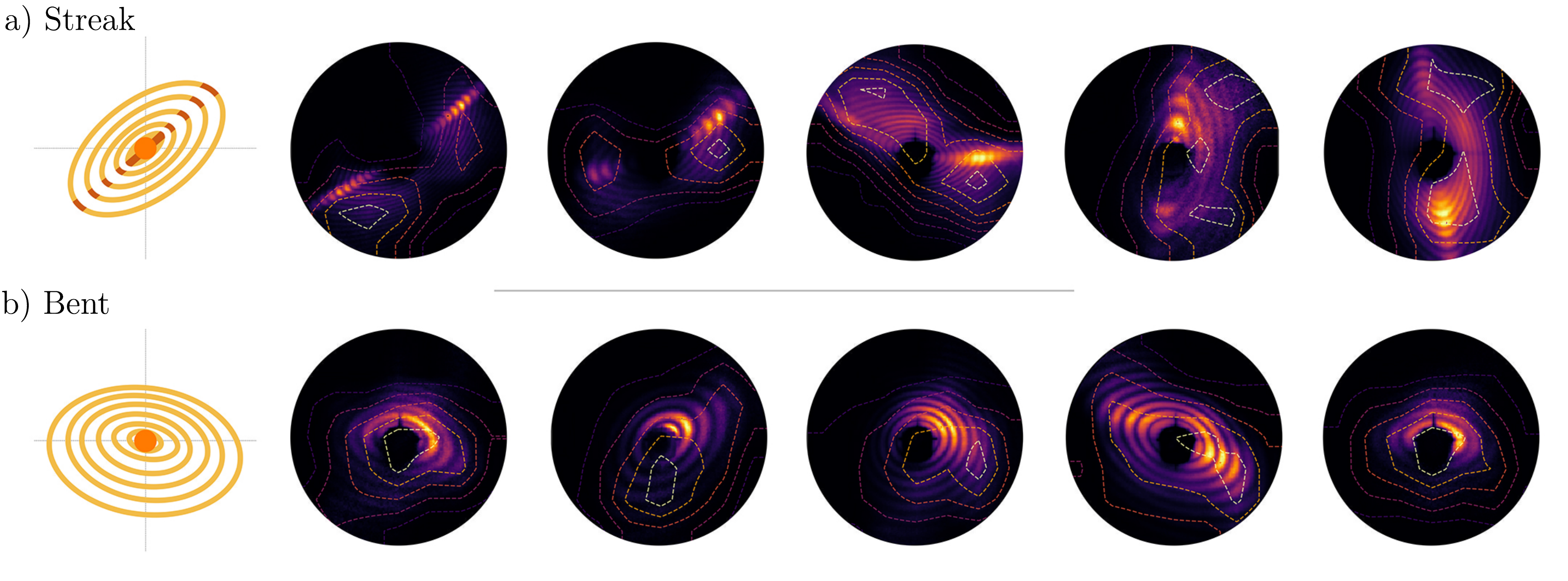}%
 \caption{\label{fig:helium_grads} Showing the GradCam++ results for two distinct classes from the helium dataset. a) shows five randomly selected images from the \emph{Streak} class and b) shows five images from the \emph{Bent} class. We chose these classes due to their distinct and distinguishable characteristic shapes which can easily be identified using the contour maps provided by the GradCam++ algorithm. For each class, we plot the schematic from Figure \ref{fig:helium_samples} also at the beginning of each row. GradCam++ contour levels are plotted as dashed lines and used as transparency value for the images from which we calculated them. This way regions with strong gradients are also brighter.}
\end{figure*}

Neural networks are often considered being a black box approach. We usually do not impose a-priori knowledge on our model, the network learns this on its own. Although this is part of the reason why they are so successful it also gives rise to doubts about the interpretability of their predictions.
Some ways to interpret the processes of decision finding within a trained neural network have been presented in the literature \cite{Selvaraju2017,Chattopadhyay2017,Zhou2015,Li2018}. In order to get a better understanding of why our deep neural network assigned images to certain classes, we calculated heatmaps using the GradCam++ algorithm \cite{Chattopadhyay2017}. These heatmaps are making visible where the network has looked for in a particular class, which we do by tracing back the gradient flow from the output layer to the last convolutional layer. The network's class-specific interest directly correlates with this gradient signal because, in essence, we simulate a training step using backpropagation and interpolate the feature maps from the last convolutional layer. A full description of this process is given in appendix \ref{app:grad_cam}. The output of the GradCam++ algorithm provides contour maps whose amplitude is a normalized measure for how much the gradient would impose corrections on the weights if used during training. This gradient flow directly corresponds to what the network deemed the most relevant regions.

Figure \ref{fig:helium_grads} shows the GradCam++ results for the \emph{Streak} and \emph{Bent} classes using our best performing network - $\text{ResNet18}_{\alpha=0.2}$. We present results from these classes, because the distinct spatial characteristics are obvious to the human eye. Therefore they are an ideal candidate to test if the neural network understood these characteristics. In each row of Figure \ref{fig:helium_samples}, a schematic sketch of the key feature together with five randomly selected images from this class are depicted.


The GradCam++ contour maps are overlaid on the image, in addition, the contour levels are also used as an $\alpha$ mask for the diffraction image so that the brightest areas in each plot correspond to the ones with the highest gradient flow. In the case of the \emph{Streak} class, Figure \ref{fig:helium_grads} clearly shows that the neural network was able to identify the dominant streak feature regardless of its orientation or size. Results on the \emph{Bent} class also show a strong correlation between the shape of the contour maps and the bent shape of the diffraction pattern.

Therefore, combining these metrics and the GradCam++ images we think that the \emph{Streak} class feature identified by the neural network indeed corresponds to the one seen by the researcher. Also, the \emph{Bent} class contour maps from the network show a clear resemblance of the feature intended by the researcher, albeit not so strongly pronounced. Although the deep neural network learned these representations on its own, they co-align with the intentions of the researcher. This demonstrates that neural networks are capable of learning these complicated patterns on their own.

\section{\label{sec:summary}Summary and outlook}

In this paper, we give a general introduction on the capabilities of neural networks and provide results on the first domain adaption of neural networks for the use-case of diffraction images as input data. The main additions of this paper are (i) a novel activation function that incorporates the intrinsic logarithmic intensity scaling of diffraction images, (ii) an evaluation on the impact of different training set sizes on the performance of a trained network and (iii) the use of the point-wise cross-correlation function to improve the resistance against very noisy data. In addition, we provide a large benchmarking routine, utilizing multiple neural network architectures and layouts in appendix \ref{subsec:architecture_choices}.

We have shown that even in the most basic configuration, convolutional deep neural networks outperform previously established sorting algorithms by a significant margin. More importantly, we improved on these baseline results by modifying the activation function for the first layer. For the case of very noisy data, often a problem in diffraction imaging experiments, we showed that two-point cross-correlation maps as input data instead of the original diffraction images improve the robustness of the classification capabilities of the network.
Our results set the stage for using deep learning techniques as feature extractors from diffraction imaging datasets. The ultimate goal will be establishing an unsupervised routine that can categorize and extract essential pieces of information of a large set of diffraction images on its own.
We envision for the near future, that the gained insights lead to multiple new approaches regarding neural networks and diffraction data. For example, the MSFT algorithm used in \citet{Langbehn2018}, can be used as a generative module in an end-to-end unsupervised classification routine using large synthetic datasets as training data for a neural network. This approach can be extended to utilize these trained networks as an online-analysis tool during the experiments. Furthermore, we hope to develop an unsupervised approach that connects the recent research from Generative Adversarial Network theory \cite{Zhang2018a,Miyato2018,Miyato2018a,Goodfellow2016} and mutual information maximization \cite{Chen2016a} with the results of this paper. Such an approach would allow for finding characteristic classes of patterns within a data set without any a priori knowledge about the recorded data. All of the code, written in Python $3.6+$ and using the Tensorflow framework, is available at Github, free to use under the MIT License \footnote{\url{https://github.com/julian-carpenter/airynet}}. We hope the community uses and improves the code provided in this repository.



\begin{acknowledgments}
 We would like to thank K. Kolatzki, B. Senfftleben, R.M.P. Tanyag, M.J.J. Vrakking, A. Rouzée, B. Fingerhut, D. Engel and A. L\"ubcke from the Max-Born-Institute, Ruslan Kurta from the European XFEL and Christian Peltz as well as Thomas Fennel from the University of Rostock for fruitful discussions. This work received financial support by the Deutsche Forschungsgemeinschaft under Grant \emph{MO 719/13-1}, \emph{14-1} and \emph{STI 125/19-1} and by the Leibniz Grant \emph{SAW/2017/MBI4}.
\end{acknowledgments}

\bibliography{library}
\newpage
\appendix
\section{\label{subsec:architecture_choices}Architectural design choices}

In this section, we describe and explain our choices for neural network architecture to establish as baseline performance when working with diffraction patterns, before the inclusion of our diffraction specific activation function, see section \ref{subsec:activation_function} in the main manuscript. We present the theory and background on available architectures and provide results on two architectures with five depth layouts.


There are different layer styles from which we can build a neural network. Nomenclature is that a full arrangement of all layers is called architecture, or configuration, of the network.

For our tests, we use two different neural network architectures, a ResNet and a VGG- Net, both with multiple depth layouts. For the ResNet, we train and evaluate three depth variations (\num{18}, \num{50} and \num{101} layers), and for the VGG-Net we train two variants (\num{16} and \num{19} layers).


The structure of this section is as follows: First, we explain how a convolutional layer works in general. Second, we motivate the derivation of the VGG-Net from preceding architectures, and third, we show how the ResNet architecture can be explained by expanding the core ideas used in the VGG-Net. In the following section, we will then present the results  for all the here trained configurations.

Almost every architectural design is empirically derived \cite{Szegedy2016,Schmidhuber2014,LeCun2015} and constitutes of multiple combinations of only a few basic layer styles, namely the fully connected layer, convolutional layer, a pooling operation and a batch normalization operation. We discuss the pooling and batch normalization layer only in appendix \ref{app:architecture}, because of their minor role within the neural network. The reader is also referred to the exhaustive overview in \citet{Schmidhuber2014} and \citet{LeCun2015}. Since the convolutional layer serves as a fundamental basis for image analysis with neural networks, we explain it here in more detail.

The very basic idea of a convolutional layer is that nearby pixels in an input image are more strongly correlated than more distant pixels, this is called a local receptive field. Therefore, by calculating a convolution over an input image with a trainable filter of size $>1\times1$ we can approximate these correlations.

\begin{figure*}
 \includegraphics[width=1\textwidth]{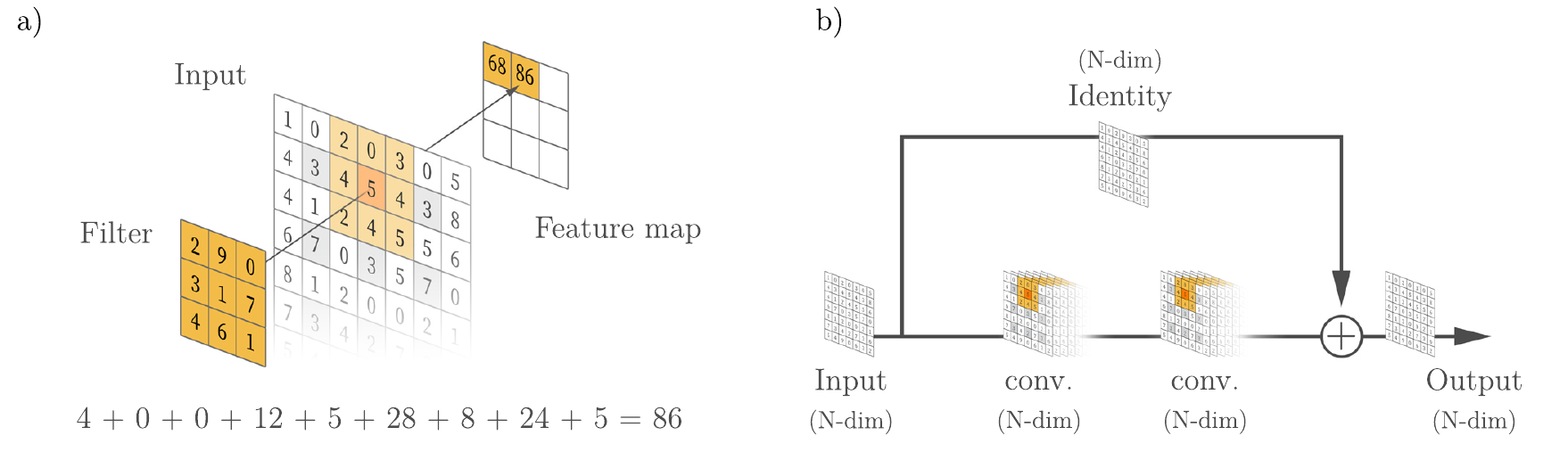}%
 \caption{\label{fig:conv_skip_layer} Schematic for a convolutional operation inside a convolutional layer in a), and for a classic skip connection found in the ResNet architecture in b). a) illustrates the local receptive fields and shared weights concept. The convolutional filter has size $3\times3$ and stride \num{2} and is sliding over the input image of size $7\times7$, which produces an output, called feature map, of size $3\times3$. The stride is the distance the filter is moving in each step which is implied by the gray shading every \num{2} pixels in the input image. Using a local receptive field describes the inclusion of nearby pixels, and weight sharing means using the same filter weights for the whole input image. The calculation at the bottom is for the second entry in the feature map. b) A classical skip connection is shown with two convolutional layers that approximate a sparse residue which gets added to the identity at the output.}
\end{figure*}

In a convolutional layer, $N$ filter, with size $M\times M$, slide over an input image and produce $N$ convolved maps, called feature maps. One filter uses the same weights on all parts of the input image for producing one feature map; this is called weight sharing. Weight sharing reduces not only the complexity of the model but provides a bridge towards the convolution function in mathematics. With weight sharing, we can identify the filter within the convolutional layer as a kernel function from the mathematical convolution function. Figure \ref{fig:conv_skip_layer} a) shows a schematic of a convolutional layer with one filter.

This exemplary filter with size $3\times3$ slides over an image of size $7\times7$ and producing a feature map of size $3\times3$. The feature map is smaller than the input image because the filter moves two pixels for each step. This step-size is called \emph{stride}.

Hereafter we use the notation \emph{conv(a, b, c)} for a convolutional layer with filter size $\text{a}\times \text{a}$, number of filters $b$ and stride $c$. The example from figure \ref{fig:conv_skip_layer} a) could, therefore, be written as $\text{conv}\left(3,~1,~2\right)$ and would result in \num{9} trainable weight parameters plus \num{1} bias parameter (not shown in the figure).


This concept was introduced with the LeNet architecture by \citet{Lecun1998} which is considered the seminal work in the field and the first deep convolutional neural network. After Yann LeCun proposed the LeNet architecture, further research \cite{Krizhevsky2012} led to the now de-facto standard for plain convolutional networks, the VGG-Net. \citet{Simonyan2014} proposed the original architecture which consists of up to \num{19} weight layers of which \num{16} are convolutional layers, and \num{3} are fully connected ones.

It is easy to build, easy to train and provides in general good results \cite{Schmidhuber2014,LeCun2015}. For these reasons, we include two variations of it in our tests, namely version D and E, nomenclature is from \cite{Simonyan2014}). Table \ref{tab:vgg_architecture} shows the details of the architecture, using the naming convention we introduced with the convolutional layer.

\citet{Simonyan2014} derived the VGG-Net directly from the LeNet by arguing that three convolutional layers with filter size \num{3} and stride \num{1} (VGG-Net) achieve better results than only one filter with size \num{7} and stride \num{2} (LeNet), which equals to the same effective local receptive field size \cite{Simonyan2014}. Three layers perform better than one due to having \num{2} additional non-linear activation functions and reduced complexity (less weight-parameter because of the smaller filter sizes), which enforces the neural network not only to be more discriminative but to find sparser solutions \cite{Simonyan2014}.

\begin{table}[tbh]
 \caption{\label{tab:vgg_architecture}
  The deep neural network architecture of the VGG variant D and E. conv(a, b, c) is a convolutional layer with filter size $\text{a}\times \text{a}$, number of filters $b$ and stride $c$. max pooling(d, e) is a max pooling layer with filter size $\text{d}\times \text{d}$ and stride $e$. Note that we changed the fully connected layer of the original architecture to a convolutional layer.
 }
 \begin{ruledtabular}
  \begin{tabular}{rcc}
   \textrm{Variant}   & D                                                                                                 & E                                          \\
   \textrm{Depth}     & 16                                                                                                & 19                                         \\
   \colrule
   \textrm{Input}     & \multicolumn{2}{c}{$2\times\text{conv}\left(3,~64,~1\right)$}                                                                                  \\
   \textrm{Pooling}   & \multicolumn{2}{c}{$\text{max pooling}\left(2,~2\right)$}                                                                                      \\
   \colrule
   \textrm{Block 1}   & \multicolumn{2}{c}{$2\times\text{conv}\left(3,~128,~1\right)$}                                                                                 \\
   \textrm{Pooling}   & \multicolumn{2}{c}{$\text{max pooling}\left(2,~2\right)$}                                                                                      \\
   \colrule
   \textrm{Block 2}   & $3\times\text{conv}\left(3,~256,~1\right)$                                                        & $4\times\text{conv}\left(3,~256,~1\right)$ \\
   \textrm{Pooling}   & \multicolumn{2}{c}{$\text{max pooling}\left(2,~2\right)$}                                                                                      \\
   \colrule
   \textrm{Block 3}   & $3\times\text{conv}\left(3,~512,~1\right)$                                                        & $4\times\text{conv}\left(3,~512,~1\right)$ \\
   \textrm{Pooling}   & \multicolumn{2}{c}{$\text{max pooling}\left(2,~2\right)$}                                                                                      \\
   \colrule
   \textrm{Block 4}   & $3\times\text{conv}\left(3,~512,~1\right)$                                                        & $4\times\text{conv}\left(3,~512,~1\right)$ \\
   \textrm{Pooling}   & \multicolumn{2}{c}{$\text{max pooling}\left(2,~2\right)$}                                                                                      \\
   \colrule
   \textrm{Out block} & \multicolumn{2}{c}{$2\times\text{conv}\left(7,~4096,~1\right),~ \text{conv}\left(1,~N,~1\right)$}                                              \\
  \end{tabular}
 \end{ruledtabular}
\end{table}

Building on the results achieved by the VGG-net, it was shown that the depth of a deep neural network directly relates to its classification capabilities \cite{Eldan2015, He2015, Szegedy2014}. This led to the introduction of the so-called residual skip-connections which further exploit this \emph{depth-matters} concept \cite{He2015, Szegedy2014}. These residual skip connections are the name-giving components for the ResNet architecture.

In principle, a ResNet still uses the VGG architectural layout but exchanges the convolutional blocks \num{1} to \num{4} with residual skip connections, compare tables \ref{tab:vgg_architecture} and \ref{tab:resnet_architecture}. This exchange drastically reduces the complexity of the whole network while increasing the number of layers.

The VGG-architecture can be broken down into six blocks, one input block, one output block, and four convolutional blocks (see table \ref{tab:vgg_architecture}). Block \num{2} is the first block in which there are distinctions between VGG variant D and E.

The VGG-net architecture proved that increasing the depth and decreasing the amount and size of the filters increases the accuracy, which ultimately gave rise to the \emph{plain skip connections}: Blocks of few convolutional layers designed to replace the large amounts of filters in one layer for multiple layers with fewer, and smaller, filters. Two types exist: A classical and a bottleneck skip connection, both differ only in the amount of how much the depth is increased and the complexity decreased.

This addition has so far only modified the depth and complexity of the network and is called a \emph{plain} network, see \citet{He2015}. It performs reasonably well but not significantly better than VGG-net. A residual skip connection differs from a plain skip connection only in adding the identity of its inputs to its outputs. This way all the convolutional layers in a skip connection learn only a \emph{residual} of their input. This simple technique enables a ResNet to outperform all other convolutional deep neural network architectures \cite{Szegedy2016,He2016a}. Figure \ref{fig:conv_skip_layer} b) exemplifies a classical residual skip connection. There is still an ongoing debate about why a residual neural network performs so well  \cite{He2015, Szegedy2014, Veit2016}. Research has shown that ResNets find sparser solutions faster due to their layout, and that they behave like ensembles of shallower networks with information flow only activated on \num{10} to \num{34} layers even when the neural network has a depth of \num{101} layers \cite{He2015, Szegedy2014, Veit2016}.

However, besides empirical success, one of the critical advantages of ResNets is that reaching training convergence is not getting significantly harder when increasing the depth of the neural network, which is usually the case with other architectures. Therefore, the training of very deep residual neural networks is no more difficult than training shallow plain neural networks \cite{Li2016, Szegedy2016}.

For these reasons, we train three variants, with \num{18}, \num{50} and \num{101} layers, of a further optimized version of the classical ResNet, called pre-activated ResNet \cite{He2016a} (see table \ref{tab:resnet_architecture} for implementation details).

\begin{table*}[htb]
 \caption{\label{tab:resnet_architecture}
  Used ResNet variants, see also \num{18}, \num{50} and \num{101} layer layout in \cite{He2015}. Note that we added the pre-activated layer layout from \cite{Zagoruyko2016}. conv(a, b, c) is a convolutional layer with filter size $\text{a}\times \text{a}$, number of filters $b$ and stride $c$. \emph{max pooling(d, e)} is a max pooling layer with filter size $\text{d}\times \text{d}$ and stride e. \emph{avg pooling} is a global average pooling layer, and \emph{fc(f)} is a fully connected layer with output size f. Layers in bold emphasis have a stride of \num{2} during their first iteration, therefore reducing the dimension by a factor of \num{2}.}
 \begin{ruledtabular}
  \begin{tabular}{@{}rccc@{}}
   Variant                                            & Classic                                            & Bottleneck & Bottleneck \\
   Depth                                              & 18                                                 & 50         & 101        \\
   \colrule
   Input                                              & \multicolumn{3}{c}{conv(7, 64, 2)}                                           \\
   Pooling                                            & \multicolumn{3}{c}{max pooling(3 , 2)}                                       \\
   \colrule
   Block 1                                            & $2\times \left[\begin{array}{c}
      \text{\textbf{conv}}\left(3,~64,~1\right) \\ \text{conv}\left(3,~64,~1\right)
     \end{array} \right]$  &
   $3\times \left[\begin{array}{c}
      \text{\textbf{conv}}\left(1,~64,~1\right) \\ \text{conv}\left(3,~64,~3\right)\\\text{conv}\left(1,~256,~1\right)
     \end{array} \right]$  &
   $3\times \left[\begin{array}{c}
      \text{\textbf{conv}}\left(1,~64,~1\right) \\ \text{conv}\left(3,~64,~3\right)\\\text{conv}\left(1,~256,~1\right)
     \end{array} \right]$                                                                                 \\
   \colrule
   Block 2                                            & $2\times \left[\begin{array}{c}
      \text{\textbf{conv}}\left(3,~128,~1\right) \\ \text{conv}\left(3,~128,~1\right)
     \end{array} \right]$  &
   $4\times \left[\begin{array}{c}
      \text{\textbf{conv}}\left(1,~128,~1\right) \\ \text{conv}\left(3,~128,~3\right)\\\text{conv}\left(1,~512,~1\right)
     \end{array} \right]$  &
   $8\times \left[\begin{array}{c}
      \text{\textbf{conv}}\left(1,~128,~1\right) \\ \text{conv}\left(3,~128,~3\right)\\\text{conv}\left(1,~512,~1\right)
     \end{array} \right]$                                                                                 \\
   \colrule
   Block 3                                            & $2\times \left[\begin{array}{c}
      \text{\textbf{conv}}\left(3,~256,~1\right) \\ \text{conv}\left(3,~256,~1\right)
     \end{array} \right]$  &
   $6\times \left[\begin{array}{c}
      \text{\textbf{conv}}\left(1,~256,~1\right) \\ \text{conv}\left(3,~256,~3\right)\\\text{conv}\left(1,~1024,~1\right)
     \end{array} \right]$ &
   $36\times \left[\begin{array}{c}
      \text{\textbf{conv}}\left(1,~256,~1\right) \\ \text{conv}\left(3,~256,~3\right)\\\text{conv}\left(1,~1024,~1\right)
     \end{array} \right]$                                                                               \\
   \colrule
   Block 4                                            & $2\times \left[\begin{array}{c}
      \text{\textbf{conv}}\left(3,~512,~1\right) \\ \text{conv}\left(3,~512,~1\right)
     \end{array} \right]$ &
   $3\times \left[\begin{array}{c}
      \text{\textbf{conv}}\left(1,~512,~1\right) \\ \text{conv}\left(3,~512,~3\right)\\\text{conv}\left(1,~2048,~1\right)
     \end{array} \right]$ &
   $3\times \left[\begin{array}{c}
      \text{\textbf{conv}}\left(1,~512,~1\right) \\ \text{conv}\left(3,~512,~3\right)\\\text{conv}\left(1,~2048,~1\right)
     \end{array} \right]$                                                                                \\
   \colrule
   Output block                                       & \multicolumn{3}{c}{avg pooling, fc(N)}
  \end{tabular}
 \end{ruledtabular}
\end{table*}


Table \ref{tab:precision_recall_overall_metrics_architectures} shows the overall evaluation metrics on the helium and the CXIDB dataset. Table \ref{tab:precision_recall_overall_metrics_architectures_per_class} shows the per-class evaluation metrics for the helium dataset, which are not needed for the CXIDB dataset because predictions on the helium dataset are a multi-class problem whereas predictions on the CXIDB data are single-class. Single-class - or \emph{one-hot} - problems have identical overall- and per-class-evaluation metrics. We trained all models as described in section 3.1.5 in the main manuscript.

Table \ref{tab:precision_recall_overall_metrics_architectures} shows the overall evaluation metrics as well as the training wall time. $\text{Train Time}_{Max}$ is the time when the neural network achieved the highest accuracy score on the evaluation dataset, and $\text{Train Time}_{Full}$ is the time for training 200 epochs. However, in practice we achieved optimal convergence after training for 70 to 100 epochs. After this the network showed overfitting.

Both VGG models took significantly longer to train than the ResNet variants, needing between \SIrange{6.5}{6.7}{\hour}, for \num{200} epochs on both datasets, whereas training ResNet101 took only \SI{2.8}{\hour} and \SI{2.6}{\hour} respectively. Furthermore, the maximum reached accuracy of both VGG networks is more than half a percentage point below the maximum of ResNet101 - \num{0.959} compared to \num{0.964} for the helium data and \num{0.970} vs. \num{0.978} for the CXIDB data. Also, accuracy did not change much when increasing the depth from \numrange{16}{19}, precision even decreased slightly and recall remained unchanged.


On the other hand, increasing complexity within the ResNet architecture helped to boost the accuracy from \num{0.955} (CXIDB data: \num{0.973}) with ResNet18 to \num{0.964} (CXIDB data: \num{0.978}) with ResNet101.

\begin{table}[htb]
 \caption{\label{tab:precision_recall_overall_metrics_architectures}
  Overall evaluation metrics for all architectures and both datasets. The training time after which the neural network scored the highest accuracy score on the evaluation dataset is labeled $\text{Train Time}_{Max}$ and $\text{Train Time}_{Full}$ is the time for training the full \num{200} epochs. The table gives the max values during training for accuracy, precision, and recall. Bold scores are the best results in their respective category.}
 \begin{ruledtabular}
  \begin{tabular}{rccccc}
   Architecture                   & \multicolumn{3}{c}{ResNet} & \multicolumn{2}{c}{VGG}                                            \\
   Depth                          & \num{18}                   & \num{50}                & \num{101}    & \num{16}    & \num{19}    \\
   \colrule
   \multicolumn{6}{c}{}                                                                                                             \\
   Dataset                        & \multicolumn{5}{c}{Helium}                                                                      \\
   Accuracy                       & \num{0.955}                & \num{0.958}             & $\bm{0.964}$ & \num{0.958} & \num{0.959} \\
   Precision                      & \num{0.918}                & \num{0.917}             & $\bm{0.925}$ & \num{0.923} & \num{0.920} \\
   Recall                         & \num{0.866}                & \num{0.864}             & $\bm{0.878}$ & \num{0.867} & \num{0.867} \\
   $\text{Train Time}_{Max}$ [h]  & $\bm{0.231}$               & \num{0.605}             & \num{0.940}  & \num{3.271} & \num{6.615} \\
   $\text{Train Time}_{Full}$ [h] & $\bm{0.694}$               & \num{1.814}             & \num{2.820}  & \num{6.541} & \num{6.726} \\
   \multicolumn{6}{c}{}                                                                                                             \\
   Dataset                        & \multicolumn{5}{c}{CXIDB}                                                                       \\
   Accuracy                       & \num{0.967}                & \num{0.973}             & $\bm{0.978}$ & \num{0.970} & \num{0.970} \\
   Precision                      & \num{0.932}                & \num{0.937}             & $\bm{0.949}$ & \num{0.944} & \num{0.943} \\
   Recall                         & \num{0.933}                & \num{0.937}             & $\bm{0.941}$ & \num{0.904} & \num{0.904} \\
   $\text{Train Time}_{Max}$ [h]  & $\bm{0.278}$               & \num{1.205}             & \num{1.093}  & \num{4.374} & \num{4.480} \\
   $\text{Train Time}_{Full}$ [h] & $\bm{0.668}$               & \num{1.807}             & \num{2.623}  & \num{6.562} & \num{6.720}
  \end{tabular}
 \end{ruledtabular}
\end{table}

For the multi-class results in Table \ref{tab:precision_recall_overall_metrics_architectures_per_class}, we chose the ResNet101 layout, as it is our best performing configuration. For classes \emph{Oblate}, \emph{Spherical}, \emph{Streak} and \emph{Empty} a precision of \numrange{0.9247}{0.9770} and a recall of at least \num{0.9763} show that the majority of all predictions in these classes were correct and virtually no image was missed.

For the classes \emph{Prolate}, \emph{Bent}, \emph{Double Rings} and \emph{Layered}, the ResNet reached a good precision, but a recall score of $\approx 0.65$ shows that it missed almost a third of all available images, indicating we failed to generalize the network for these classes.

For \emph{Elliptical}, \emph{Newton Rings} and \emph{Asymmetric} images, the recall of \numrange{0.2207}{0.4836} shows that these images were a lot harder to find, observable in the relatively low precision scores for those classes. \emph{Elliptical} is the only class of these three where precision is high enough for using the neural network as a predictor. For the \emph{Newton Rings} and \emph{Asymmetric} class, with precision scores around \num{0.6}, the neural network is effectively guessing.

\begin{table}[tbh]
 \caption{\label{tab:precision_recall_overall_metrics_architectures_per_class}
  Per-class accuracy, precision and recall values for the best performing ResNet configuration with \num{101} layers. Samples are the number of images whose ground truth label is positive in the evaluation dataset. Results are shown for both datasets and reflect the maximum achieved value reached throughout the training process, assessed on the evaluation dataset.}
 \begin{ruledtabular}
  \begin{tabular}{rccccc}
   Class        & Accuracy & Precision & Recall & Samples \\
   \colrule
   Oblate       & 0.9681   & 0.9770    & 0.9965 & 988     \\
   Spherical    & 0.9166   & 0.9247    & 0.9849 & 869     \\
   Elliptical   & 0.9231   & 0.8054    & 0.4836 & 119     \\
   Newton rings & 0.9352   & 0.6325    & 0.2282 & 69      \\
   Prolate      & 0.9690   & 0.9274    & 0.6777 & 68      \\
   Bent         & 0.9657   & 0.8161    & 0.6487 & 59      \\
   Asymmetric   & 0.9458   & 0.6044    & 0.2207 & 55      \\
   Streak       & 0.9898   & 0.9372    & 0.9876 & 36      \\
   Double Rings & 0.9768   & 0.7708    & 0.6788 & 33      \\
   Layered      & 0.9896   & 0.9062    & 0.6170 & 7       \\
   Empty        & 0.9904   & 0.9537    & 0.9763 & 32      \\
  \end{tabular}
 \end{ruledtabular}
\end{table}


The performance of all variants clearly shows the general good classification capabilities of a convolutional deep neural networks in the use case of diffraction patterns. Even the lowest performing neural network can outperform previous classification approaches by a large margin - compare with \cite{Bobkov2015}. In particular, the results of ResNet18 are compelling; it is small, easy to train and has relatively low complexity. Although having only a fraction of trainable parameters, it performed almost always on-par with the much more complex VGG architectures and all this while taking only \SI{0.2}{\hour} for reaching the maximum accuracy during training.

Therefore, we chose the ResNet18 layout as the default configuration for all the following experiments; it is an ideal compromise between complexity, training time and classification accuracy.

\section{\label{app:derive_cross_entropy}Derivation of the binary cross-entropy}
Here, we give an derivation for the binary cross entropy (Equation 6 in the main manuscript). We start with the most general form of the cross-entropy given by:
\begin{gather}
 H(p, q) = H(p) + D_{\mathrm{KL}}(p \| q)
 \label{eq:cross_entropy_first_supp}
\end{gather}
where $H(p)$ is the Shannon entropy of $p$, and $D_{\mathrm{KL}}(p \| q)$ is
the Kullback–Leibler divergence of $p$ and $q$ \cite{Goodfellow2016a}. This is equivalent to:
\begin{gather}
 H(p,q) = -\sum_i p_i\log q_i,
 \label{eq:app_cross_entropy}
\end{gather}
where $p_i$ and $q_i$ are two probability distributions over the same set of events.
$p_i$ is the ``correct'' distribution, and $q_i$ is the approximation of $p_i$ from the deep neural network.
Since we are using a Bernoulli distribution as our probabilistic model there are only
two outcomes that one event ($k$) can have: $k\in\left\{0, 1\right\}$. The probability
for both outcomes of one event and of both distributions can be written as:
\begin{gather}
 p\left(x\right) =
 \begin{cases}
  y\left(x\right)   & \text{if $k=1$} \\
  1-y\left(x\right) & \text{if $k=0$}
 \end{cases} \nonumber \\
 q\left(x\right) =
 \begin{cases}
  \hat{y}\left(x\right)   & \text{if $k=1$} \\
  1-\hat{y}\left(x\right) & \text{if $k=0$}
 \end{cases} \nonumber
\end{gather}
$x$ is some event, $y$ is the ground truth label and $\hat{y}$ is the approximate probability
assigned by the deep neural network. Since we are using a sigmoid function at the output of our deep neural network, we can
simplify equation \ref{eq:app_cross_entropy}. Using:
\begin{gather}
 \hat{y}\left(x\right)_{\text{sigmoid}} = \frac{1}{1+\exp{\left(-x\right)}}, \nonumber
\end{gather}
we can write:
\begin{gather}
 \begin{align*}
  H(p,q) & = -\sum_i^2 p_i\log q_i,                                                                                                               \\
         & = - y\left(x\right) \log{\left(\hat{y}\left(x\right)\right)}-\left(1-y\left(x\right)\right) \log{\left(1-\hat{y}\left(x\right)\right)} \\
         & = - y\left(x\right) \log{\left( \frac{1}{1+\exp{\left(-x\right)}}\right)}-\left(1-y\left(x\right)\right)                               \\
         & \qquad\log{\left(1- \frac{1}{1+\exp{\left(-x\right)}}\right)}                                                                          \\
         & \setbox0\hbox{=}\mathrel{\makebox[\wd0]{\hfil\vdots\hfil}}                                                                             \\
  ~      & =x-x~y\left(x\right)+\log{\left(1+\exp{\left(-x\right)}\right)}
 \end{align*}
\end{gather}
where $x$ is an event (e.g. the activation in the output layer of the deep neural network) and $y$ is
the real label of this event.

\section{\label{app:architecture}Further building blocks of deep neural networks}

%
This section describes the \emph{pooling} layer and the \emph{batch normalization} layer in more detail. Since these components are not critical for the neural network their explanation is only here in supplemental material.

\subsection{Pooling}
There are two commonly used variants of pooling layers, the max pool, and the average pool.
The idea is to reduce the dimensionality of the output from a preceding layer
($\dim{\bm{x}}=\left(N\times X\right)$) by letting a filter,
with size $a\times a$ slide over parts of the image with step size $b$, called stride, and
let them perform a down-sample operation.

A max pool filter only takes the maximum value, and a avg pool filter
averages over all values, within its perceptive field \cite{Lecun1998,Krizhevsky2012}, this process is equivalent to a convolutional operation but instead of a matrix multiplication with a convolutional kernel the pooling operation is carried out.

\subsection{Batch Normalization}
Every layer within a deep neural network is to some point modeling the probability distribution given to it by
its preceding layer. It is a hierarchical regression problem, which becomes harder if one layer
changes key characteristics of the modeled probability distribution (e.g. the mean, variance
or the kurtosis). This shift is then further multiplied in every succeeding layer and is
therefore dependent on the depth of the network. This phenomenon is called a covariate shift
\cite{Shimodaira2000}. Although this problem is solved in a deep neural network via domain adaptation, the costs
of a covariate shift are usually much longer training times and reduced accuracy
\cite{Jiang2008}.

For this reason a batch normalization layer (bn) is used to shift the mean of the mini-batch input
to zero and to set the variance to one. This significantly reduces
the amount of training time and increases accuracy  \cite{Ioffe2015}.
\emph{bn} consists of 4 steps after which a normalized mini-batch is returned:
\begin{enumerate}
 \item Calculated the mini-batch mean:
       \begin{gather}
        \mu_{\text{mb}} = \frac{1}{m}\sum^{m}_{i=1}x_{i} \nonumber
       \end{gather}
 \item Calculated the mini-batch variance:
       \begin{gather}
        \sigma^{2}_{\text{mb}} = \frac{1}{m}\sum^{m}_{i=1}\left(x_{i}-\mu_{\text{mb}}\right)^{2} \nonumber
       \end{gather}
 \item Normalize:
       \begin{gather}
        \hat{x_{i}} = \frac{x_{i}-\mu_{\text{mb}}}{\sqrt{\sigma^{2}_{\text{mb}}+\epsilon}} \nonumber
       \end{gather}
 \item Scale and shift according to adjustable parameter:
       \begin{gather}
        y_{i} = \gamma \hat{x_{i}} + \beta \nonumber
       \end{gather}
\end{enumerate}
where $y_{i}$ is the normalized output of input $x_{i}$ and $\gamma$ and $\beta$ are adjustable
parameter.

\section{\label{app:grad_cam}GradCam++}
In chapter \num{6} of the main manuscript we show what the neural network deemed the most relevant areas within an input image. We calculated these so-called heatmaps with an algorithm called GradCam++. The main idea is based on Cam \cite{Zhou2015} and Gradcam \cite{Selvaraju2017} and allows for a very intuitive explanation for the decisions made by a convolutional deep neural network \cite{Chattopadhyay2017}.


The core principle is that the output of a convolutional deep neural network can be expressed as a linear combination of
the globally average pooled feature maps of the last convolutional layer.

\begin{gather*}
 Y^{c} = \sum_{k} w^{c}_{k} \sum_{i} \sum_{j} A_{ij}^{k}
\end{gather*}

where $A_{ij}^{k}$ is one feature map of all $k$ maps from the last convolutional layer and $w^{c}_{k}$ are the weights for a particular class prediction $c$ of feature map $k$. $Y^{c}$ is the predicted probability that
the input image belongs to this certain class $c$. In the GradCam++ formalism the weights can
be calculated:

\begin{gather}
 w^{c}_{k} = \sum_{i} \sum_{j} a_{ij}^{kc}~\text{LeakyReLu}\left(\frac{\partial Y^{c}}{\partial  A_{ij}^{k}}\right).
 \label{eq:grad_weights}
\end{gather}
where $a_{ij}^{kc}$ are the gradient weights and $\text{LeakyReLu}\left(\cdot\right)$ is
a rectified linear unit activation function, very similar to the one we used throughout the main manuscript.
$a_{ij}^{kc}$ depends only on $A_{ij}^{k}$ and $Y^{c}$ via:

\begin{gather*}
 a_{ij}^{kc} = \frac{\frac{\partial^{2}Y^{c}}{\left(\partial A_{ij}^{k}\right)^{2}}}{2 \frac{\partial^{2}Y^{c}}{\left(\partial A_{ij}^{k}\right)^{2}} + \sum_{a} \sum_{b} A_{ab}^{k} \frac{\partial^{3}Y^{c}}{\left(\partial A_{ij}^{k}\right)^{3}}}
\end{gather*}

The final heatmap, often called saliency map, can then be obtained:

\begin{gather}
 L^{c}_{ij} = \text{LeakyReLu} \left(\sum_{k} w^{c} A^{k}_{ij}\right).
 \label{eq:grad_map}
\end{gather}

So the algorithm propagates an image forward through the network, then calculates the gradients
until the last convolutional layer, and using equation \ref{eq:grad_weights} and \ref{eq:grad_map}, obtains a heatmap of the areas within the input image that shows the gradient flow from the convolutional layer.
\end{document}